\documentclass[12pt]{article}
\usepackage{graphicx}
\begin{document}
\title{Connecting phase transitions between the 3-d O(4) Heisenberg model 
and 4-d SU(2) lattice gauge theory}
\author{Michael Grady\\
Department of Physics\\ State University of New York at Fredonia\\
Fredonia NY 14063 USA\\Ph:(716)672-6697 Fax:(716)673-3347\\grady@fredonia.edu}
\date{\today}
\maketitle
\thispagestyle{empty}
\begin{abstract}
SU(2) lattice gauge theory is extended to a larger coupling space where the coupling parameter for horizontal (spacelike)
plaquettes, $\beta _H$, differs from that for vertical (Euclidean timelike)  plaquettes, $\beta _V$.  When
$\beta _H \rightarrow \infty$ the system, when in Coulomb Gauge, splits into multiple independent 3-d O(4) Heisenberg models on spacelike 
hyperlayers. Through consideration of the robustness of the Heisenberg model phase transition to small perturbations, 
and illustrated by Monte Carlo simulations, it is shown that the ferromagnetic phase transition in this model
persists for $\beta _H < \infty$.  Once it has entered  the phase-plane it must continue to another edge due to its 
symmetry-breaking nature, and therefore must necessarily cross the $\beta _V = \beta _H$ line at a finite value.  Indeed, a higher-order
SU(2) phase transition is found  at $\beta = 3.18 \pm 0.08$, from a finite-size scaling analysis of the 
Coulomb gauge magnetization from
Monte Carlo simulations, 
which also yields critical exponents. An important technical breakthrough is the use of open boundary conditions, which
is shown to reduce systematic and random errors of the overrelaxation gauge-fixing algorithm by a factor of several hundred.
The string tension and specific heat are also shown to be consistent with finite-order scaling about this critical point 
using the same critical exponents.
\end{abstract}

\section{Introduction}
It would be very nice to have a local order parameter for confinement. On large lattices the Polyakov loop
becomes extremely small in both phases, as do large Wilson loops. In addition, it is difficult to accurately
split off the area-law confining term of Wilson loops from the other terms present. For spin models with
spontaneously broken symmetries it is much easier to study the ferromagnetic phase transition, due to the local magnetic
order parameter which (along with its moments) is easy to measure on any size lattice.

Indeed, there actually is such a local order parameter, the ferromagnetic transition of which signals deconfinement.
It requires, however, fixing configurations to minimal Coulomb gauge. Since this is an iterative gauge-fixing procedure,
it is fairly costly, but the benefit of a local order parameter may be worth it.  Nearly two decades ago, it was envisioned
that this order parameter could become a prime means for studying confinement\cite{mp}. Difficulties in gauge-fixing have
largely prevented these techniques from becoming mainstream.

The minimal Coulomb gauge involves maximizing the trace of all links in three of the four lattice directions (these
three are called ``horizontal" directions below). The fourth-direction (vertical) links can be reinterpreted as O(4) ``spins," the local 
order parameter. The minimal Coulomb gauge has a set of remnant SU(2) symmetries, global on the three horizontal-directions, 
but still local along the fourth. This is because if links are written as $a_0 + i \vec{a} \cdot \vec{\tau}$ the $a_0$ components
of the {\em horizontal} links are invariant under such a gauge transformation, and this is the quantity being maximized to 
satisfy the gauge condition.  Separate SU(2) transformations on the two ends of the fourth-direction links, which are
being treated as spins, generate for them a global SU(2)$\times$SU(2) = O(4) symmetry, which breaks spontaneously 
if the spins magnetize. In the
limit $\beta \rightarrow \infty$, the gauge condition sets the horizontal links to unity, and the plaquette interaction
collapses into a spin interaction, a dot-product between the O(4) spins.  Thus, in this limit, the SU(2) lattice
gauge theory becomes a set of 3-d O(4) Heisenberg models (or non-linear sigma models) at zero 
temperature (using $T=1/\beta$ as temperature). 
As $\beta$ becomes finite,
then the primary interaction is still the Heisenberg one, since the horizontal links still are on average close to 
the identity, but the deviations from the identity introduce new interactions between the spins, which vary spatially.
For large $\beta$ these interactions are either small or, if large, rare. There are no direct interactions between
vertical links in different hyperlayers; they interact indirectly through shared horizontal links. For a fixed set of horizontal links, the spin-model now
has the appearance of a spin-glass, with the deviations of horizontal links from the identity introducing spatially-varying
``disorder interactions."  As $\beta$ is lowered, both the pseodo-temperature, $T=1/\beta$, and the amount of disorder
from the deviation of horizontal links from the identity increases. Thus, one expects that the ferromagnetic phase that
must exist at $\beta = \infty$ (since the 3-d O(4) spin model is ferromagnetic at zero temperature), should either have
a transition to a paramagnetic phase, or possibly first to a spin-glass ordered non-ferromagnetic phase followed by
a second transition to a paramagnetic phase.  Because the strong coupling limit is completely random, the system must 
eventually
enter a paramagnetic phase. 

Conventional thinking would have the transition to paramagnetic phase happen immediately
at $T = 0+$, putting the entire Wilson axis except $\beta = \infty$ in the paramagnetic phase. However, ferromagnetism in
the Heisenberg model is quite robust against the introduction of disordering interactions. For instance the $\pm J$
3-d O(3) spin-glass model remains ferromagnetic at zero temperature until 21\% of randomly 
selected interactions are changed from ferromagnetic to 
anti-ferromagnetic\cite{robust}. The O(4) model is expected to behave similarly.  
Thus it is a little hard to imagine how the infinitesimally small disordering interactions introduced
at $T=0+$ in the gauge theory can so effectively kill the ferromagnetic order. Recently, using spin-glass methods,
a spin-glass to paramagnet transition was observed at $\beta = 1.96$, strongly 
supporting the second hypothesis above\cite{su2sg}.
Unlike the ferromagnetic phase, which is necessarily unconfined \cite{zw}, the spin-glass phase, which has
a hidden pattern of frozen disorder, is still confining.  Below, the necessary spin glass to ferromagnet
phase transition is searched for. That itself could be at $T = 0+$, again consistent with
the whole $\beta$-range being confining,
however, the evidence presented below from a variety of angles points to this transition taking place
on the infinite lattice around $\beta = 3.2$. 

A previous attempt at finding this transition showed a possible transition around $\beta = 2.6$\cite{su2slgold}. 
However this
was not consistent with the apparently obvious confinement in the region $\beta=2.7$ to 2.85 from the Polyakov loop
and interquark potential\cite{pl,iqp}.  It was later discovered that the gauge fixing procedure was not working well during
Polyakov-loop tunneling events.  This depressed the magnetization on smaller lattices resulting in a premature
crossing of the Binder cumulant.  It is not surprising that the local gauge-fixing algorithm has trouble with the
global constraints imposed by the gauge-invariant Polyakov loops. 
Switching to open boundary conditions
seems to completely eliminate this pathology, and as a result not only lessens the systematic error, but vastly
reduces the random error in quantities such as the Binder cumulant (by a factor of several hundred).  
Evidence is given below that the remaining systematic 
error is at most of order, and probably much less than, this much-reduced random error, which itself is 100 times smaller than
the size of ordinary thermal fluctuations between configurations. Even with the reduction of random error that 
results from averaging over gauge configurations, the possible systematic error remains less than or of order the net random error.
Since the crossings identifying the new phase transition 
are verified at several $\beta$'s to more than 9$\sigma$, systematic error no longer appears to be an issue.  Indeed, 
the use of open boundary conditions seems to have the
potential to transform Coulomb-gauge studies to the high-precision realm.  The use of open boundary conditions in gauge theories
has recently been justified by L\"{u}scher and Shaefer\cite{luscher} and used successfully by them to reduce 
barriers to changes in the topological charge.  The negative features of open boundary conditions,
such as the lack of translational invariance, are not so important for the large lattices in use today, and
are rather easily dealt with.

In section 2 the effectiveness of the Coulomb Gauge relaxation using open boundary conditions is detailed. In section 3, the 
larger coupling space where purely horizontal plaquettes have a different coupling parameter, $\beta _H$, than those with 
a vertical link, $\beta _V$, is explored.  This plane includes both the 3-d O(4) Heisenberg model 
on the $\beta _H = \infty$ edge, and 
the ordinary 4-d SU(2) gauge theory along the $\beta _H = \beta _V$ line.  A transition very 
similar to the Heisenberg
transition in terms of critical exponents is seen for large but non-infinite $\beta _H$.  At this position the transition 
has not so high critical exponents as seen for the $\beta _V = \beta _H$ case
and is very easily studied with standard methods (Binder cumulant crossings etc.). An important point is that once
it has been established that the transition enters the ($\beta _H$,$\beta_V$) coupling plane from the Heisenberg edge, 
it must continue until
it hits another edge. Due to the symmetry-breaking nature of the transition, the coupling plane must be split into separated
symmetry-broken and unbroken regions. As $\beta _ H$ is reduced, the critical $\beta _V$ grows, thus a transition at finite
$\beta = \beta _V = \beta _H$ has to exist.  Because the large $\beta_H$ transition is so similar to the Heisenberg
transition, and so easily established with standard Monte Carlo methods, 
this argument is perhaps the most solid evidence yet that zero-temperature 
SU(2) lattice gauge theory must have a  phase transition at finite $\beta$. This transition breaks the remnant symmetry
remaining after the Coulomb gauge fixing. Such a ferromagnetic phase is known to be non-confining\cite{zw}. 
In short, from a statistical 
mechanics point of view, what is being shown here is that SU(2) lattice gauge theory, like many other lattice gauge theories,
has a lot in common with spin theories with the same symmetry once the gauge is fixed. It is essentially a magnet.  

The SU(2) transition
itself is studied with Monte-Carlo simulations in section 4.  The open boundary conditions allow for much more precise 
Coulomb-gauge data than
previously available.  Here the transition, seen at $\beta = 3.18 \pm 0.08$, is somewhat harder to study from a 
technical standpoint, because the 
critical exponents and the cumulant crossing values are fairly extreme (in comparison to the Heisenberg model). However,
even here the cumulant crossings can be verified to more than 10$\sigma$.   

Because the effective O(4) spin models are on 3-d hyperlayers, there are some interesting dimensional issues concerning
critical behavior. When these are sorted out, fits give a correlation-length exponent $\nu = 1.7$ for the 4-d
gauge theory.  This makes the corresponding singularity in the specific heat much too soft to be seen numerically.
However, the specific heat (which shows almost no finite lattice size dependence) can still be fit to the
infinite-lattice functional form associated with such a transition.  In section 5 it is found that the range of $\nu$ to
which the sharp rise of specific heat in the crossover region can be fit is consistent with this value, as
is the $\nu$ found from a similar fit to string tension measurements and from $\beta$-shift fits
to Polyakov-loop transition points for asymmetric lattices.  Thus the behavior of these important 
gauge-invariant quantities
is consistent with correlation length scaling associated with a higher order phase transition at $\beta= 3.2$,
using the critical exponent found from the Coulomb-gauge magnetization scaling behavior. The final picture that emerges
is quite unconventional.  The SU(2) lattice gauge theory appears to have three phases: a strong-coupling
paramagnetic phase from $\beta = 0$ to 1.96, a spin-glass phase from $1.96 < \beta < 3.2$ and a non-confining
ferromagnetic phase for $ \beta > 3.2$.  The existence of the spin-glass phase explains why there is a gap
(the crossover region) between regions that are well-approximated by strong and weak coupling approximations.

\section{Coulomb gauge fixing and boundary conditions}

The first thing to be mentioned here is that these are not technically ``Coulomb gauge simulations."  To actually stay
within the Coulomb gauge during Monte Carlo simulation would be difficult and would also require adding
ghost fields to take care of the Fadeev-Popov determinant.  Rather, the Coulomb gauge fixing here is part of the
post-simulation measurement procedure, applied to a completely conventional gauge-invariant Monte-Carlo simulation.
In fact, provided that the gauge fixing algorithm works well, the absolute-value of the resulting fourth-direction
link magnetization is actually gauge-invariant in the following sense. If a gauge configuration is subjected to
a random gauge transformation and then reprocessed through the Coulomb gauge fixing algorithm, the same result should
ensue, provided there is a unique minimum. The direction of the magnetization will be different due to the 
possible inclusion of a remnant global symmetry transformation, 
but all important
results are dependent only on the absolute value of the magnetization and its even moments, as usual. 
The minimal Coulomb gauge on a finite
lattice is not generally subject to the Gribov ambiguity, because the chances of not having an absolute minimum are
quite small.  However, it is also numerically difficult to {\em find} an absolute minimum and relaxation algorithms will
likely find local minima. So, whereas in principle there is a unique solution, in practice many solutions are found 
and in that practical sense there is still a ``Gribov problem" on the lattice. 

The method used here to set the Coulomb Gauge is iterative overrelaxation, with a 70\% overshoot. There are a number
of techniques which have been tried to improve on this basic algorithm.  One is to subject the initial configuration
to random gauge transformations and then relax each of these separately.  This is done, say, 10 times. 
The one which achieves the best maximization
of the traces of horizontal links is then chosen, and the others discarded.  
Another more-sophisticated approach is
simulated annealing\cite{simanneal}.  Both are quite demanding of computer time.  

The basic overrelaxation algorithm
utilizes local gauge transformations at each site, cycling through the lattice multiple times. From 40-4000 sweeps
through the lattice are needed for reasonable convergence (the larger number for lower $\beta$). However, with periodic 
boundary conditions, this does not exhaust all of the symmetry transformations available. There are still the
Z(2)$^4$ Polyakov loop transformations in which a hyperlayer of spins pointing in a particular direction
are all multiplied by -1, the non-trivial element of the center. It has been found that adding these transformations
in a preconditioning step where the best of the 16 Polyakov-loop sectors is found first, before overrelaxation, improves
the final results considerably\cite{precon}.

When one uses multiple relaxations, most give similar results, however now and then a really bad solution is found
(local minimum far from the global one). So it is likely that using only a single relaxation would result
in too many outliers; using more relaxations will reduce these but at an obvious computational cost.
The technique chosen for the current study is somewhat different. The idea is to couple
the gauge-fixing more closely with the original simulation. Gauge fixing is performed after each sweep of the
simulation, which uses a Metropolis algorithm with about 50\% acceptance. In most cases, new gauge elements are
not chosen from the full group, but from a limited neighborhood of the original link.  Throughout most of this study,
this neighborhood was chosen to be the nearest half of the group space, but the adjustability of this hit-size
also gives one a handle to test the efficacy of the gauge-fixing algorithm. The previous gauge-fixed configuration is
used as the starting point for the Monte Carlo sweep.  Thus, since only half of the links are changed, and those that do
are mostly changed by small amounts in each sweep, the gauge is still ``partially fixed" at the beginning of the next
overrelaxation. This better starting point seems to avoid the ``really bad minima" which are sometimes found when starting
from a random gauge transformation. Since the overrelaxation is done even during the initial equilibration portion
of the simulation, it has a long time to find a good ``groove" and maintain it while the gauge configuration slowly
evolves.  Although the preconditioning step mentioned above is included as well, it turns out not to work with
this close-coupled algorithm, because, since after only one sweep the memory of the previous relaxation is still there,
the algorithm always prefers not to jump to another Polyakov loop sector.  One could possibly spawn 16 new
configurations and relax each one, but that is simply too costly.

\begin{figure}\centering
                      \includegraphics[width=4in]{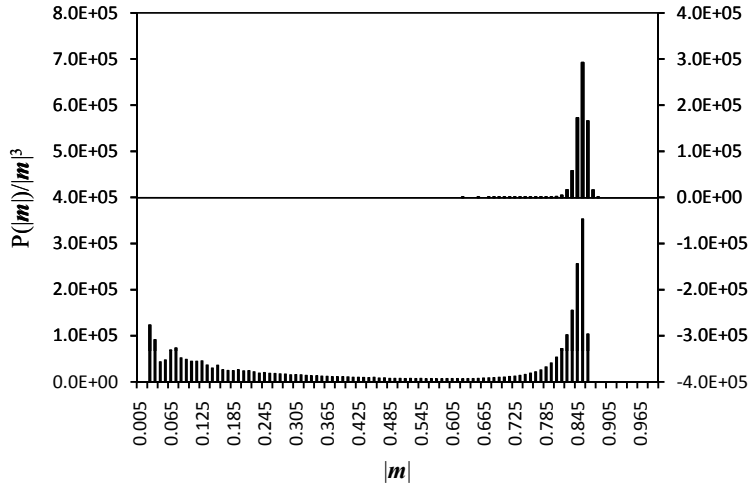}
                                 \caption{Coulomb gauge magnetization histograms at $\beta =3.5$ 
on $16^4$ lattices for periodic boundary conditions (lower)
and open boundary conditions (upper).  The lower peak on the lower graph contains 
very few actual configurations, bars 
being magnified by the $1/|\vec{m}|^3$ geometry factor. Left scale is for lower graph and right for upper.}
          \label{fig1}
       \end{figure}
This algorithm seems to perform well most of the time, but it was noticed that at high-beta ($\beta > 2.9$), where
the magnetization was quite large and histograms peaked very far from zero, occasional ``storms" of low
values occurred. This produced a small secondary peak about zero magnetization (Fig.~1-lower). 
At first this was possibly thought to be a real phenomenon, and evidence for a first-order phase transition, but 
the persistence of this second peak over a large range of $\beta$ was not consistent with that interpretation.
It was later discovered that the observed ``storms" of low magnetization appeared to occur
 during Polyakov loop tunneling events.
The known sensitivity of the gauge-fixing algorithm to Polyakov loop sector changes pointed to
it as the cause of the storm - it was simply having trouble finding a good minimum during the tunneling. Since
this represents a global shift in the vacuum, akin to a quantum phase transition, it is not surprising that
the local gauge-fixing algorithm has trouble initially following it. After a few tens of sweeps it finds a new groove,
but even a few percent of bad sweeps with very different values will affect moments of the magnetization
unacceptably.  The memory effect of the close-coupled algorithm is a negative feature when it comes to following
sudden vacuum tunnelings.  These problems only appear to occur at values of $\beta$ above the point at which the Polyakov loop
is broken on a given lattice (e.g. $\beta > 2.7$ on the $16^4$ lattice).

In order to test this further, the algorithm was tried with open boundary conditions (OBC).  
Results were beyond all expectations.
The magnetization distribution completely loses the suspicious second peak (Fig.~1 - upper graph), and 
the gauge condition function
no longer has large excursions. These reduced fluctuations also resulted in much lower statistical errors on
quantities such as susceptibility and Binder cumulant.  This suggests that much of the observed error in the periodic 
boundary condition (PBC) simulations
did indeed come from fluctuations in the efficacy of the gauge setting algorithm.

The magnetization order parameter is defined first by defining an O(4) vector $\vec{a} = (a_0$, $a_1$, $a_2$, $a_3$) 
from each 4th-direction pointing SU(2) link $a_0 \protect{\bf{1}} + i\sum _{j=1}^{3} a_j \tau _j$ and averaging these
over each 3-d perpendicular hyperlayer.
\begin{equation}
\vec{m} = \frac{1}{L^3}\sum _{\rm{hyperlayer}} \vec{a} .
\end{equation} 
Expectation values are then taken as ensemble averages over both hyperlayers and gauge configurations.
In Fig.~2, scatter plots are shown for multiple relaxations following random gauge transformations on the same gauge 
configuration, one taken from a $16^4$ lattice at $\beta = 3.5$. (Note it is the high $\beta$ region that is problematic in
the PBC simulations and data in this region is also the most important for this paper). The gauge function being maximized
is shown on the horizontal axis, and the magnetization $|\vec{m}|$ 
on the vertical axis with (a) showing
the PBC case and (b) the OBC case. 
\begin{figure}
                      \includegraphics[width=2.5in]{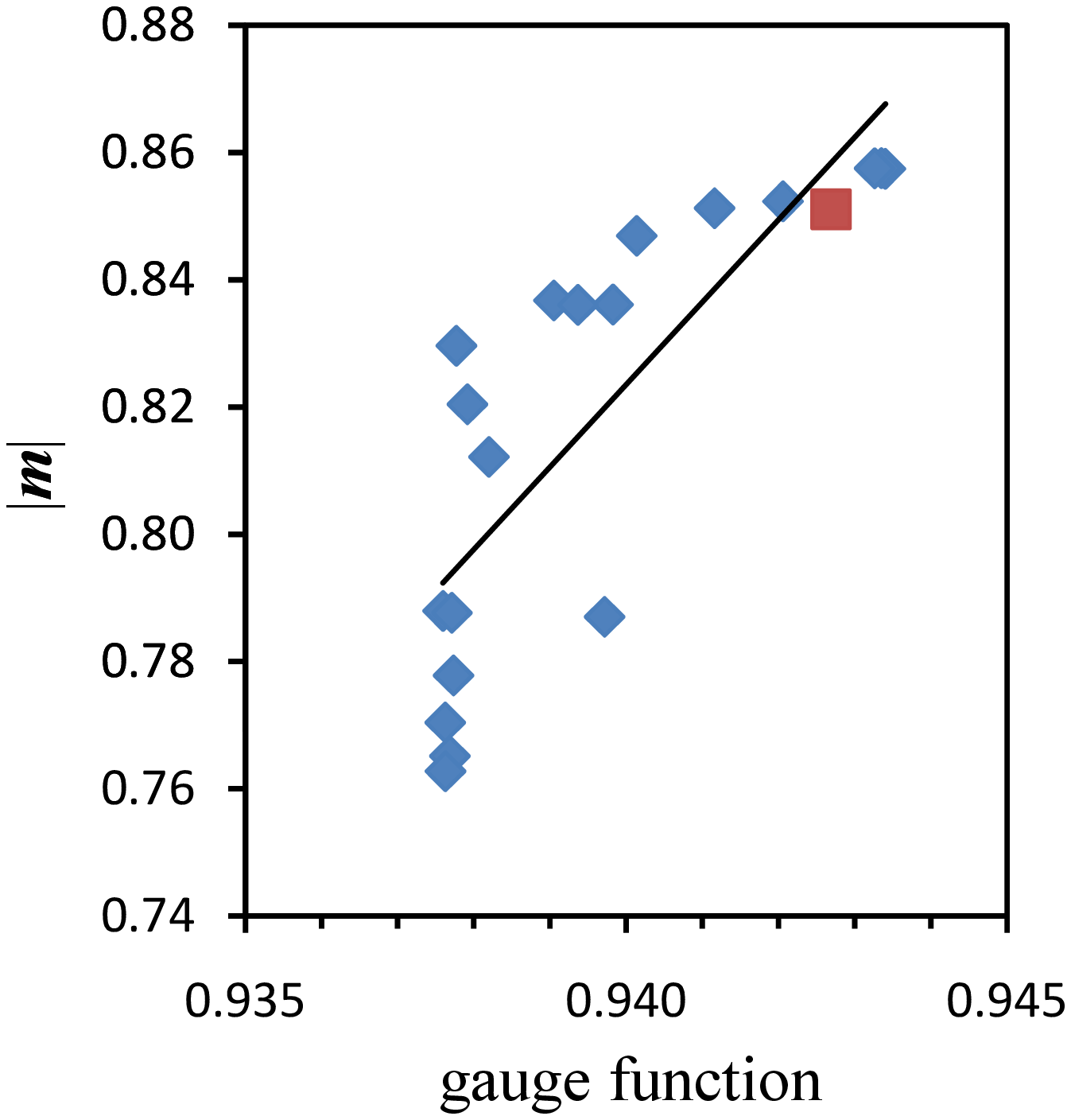}\includegraphics[width=2.5in]{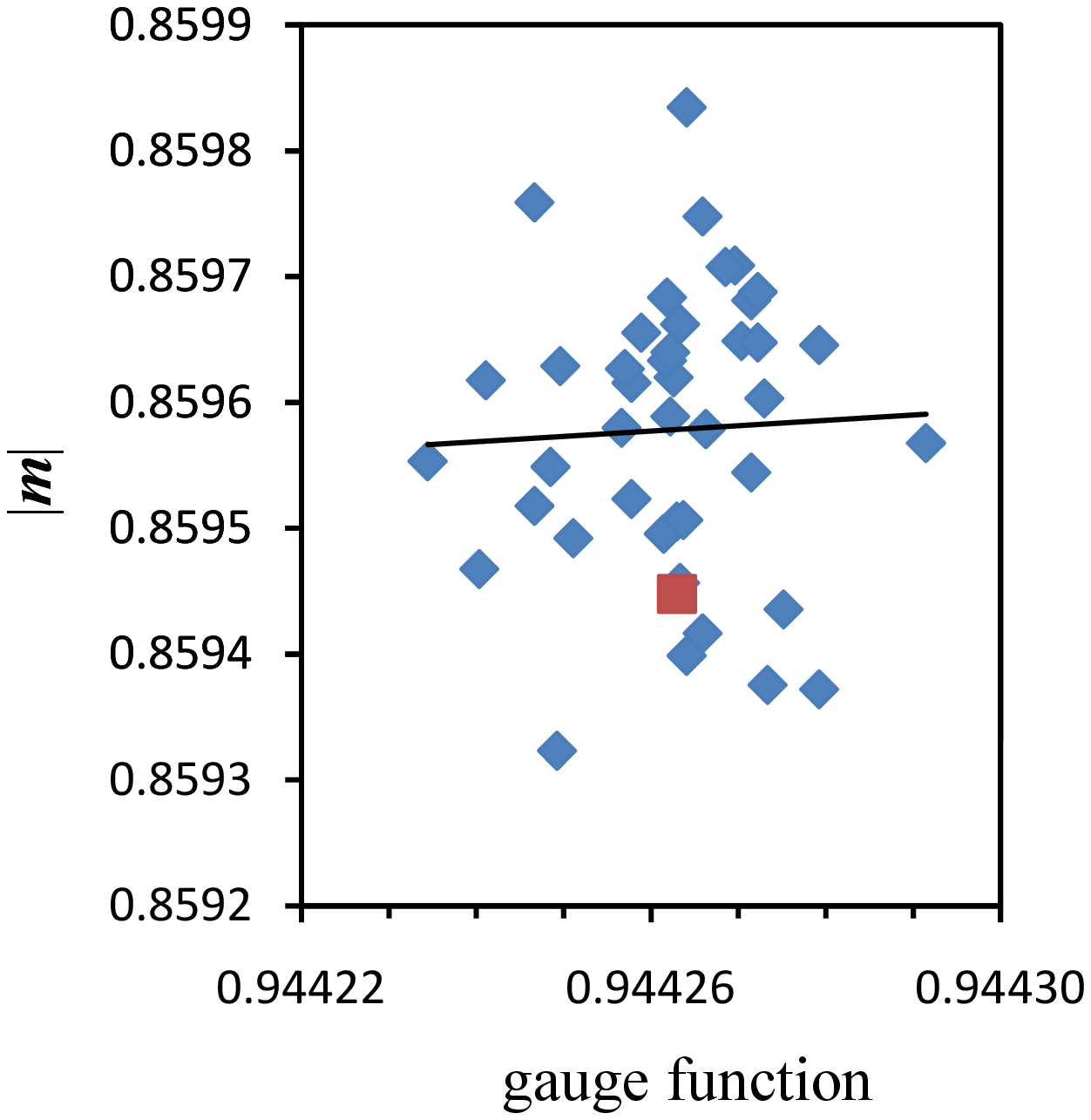}
                                 \caption{Results of multiple gauge relaxations on the same gauge configuration
after random gauge transformations for (a) periodic boundary conditions and (b) open boundary conditions.  
The single square datapoint on each graph is the result 
from the closely coupled algorithm.  Trend lines indicate the remaining degree of correlation 
between magnetization and gauge function.}
          \label{fig2}
       \end{figure}
Note the large difference in scales.  
The scatter for the OBC case is seen to be smaller
by a factor of about 200. 
Also for OBC there is very little correlation between the small remaining fluctuations in gauge function with the observable
of interest, the magnetization.  This would seem to indicate that the residual error is mostly random, rather than systematic,
and thus easily observed. This contrasts with the rather strong correlation seen with PBC, which suggests a systematic error at 
least as large as the random.  In addition, the OBC fluctuations are around 100 times smaller than the ordinary thermal fluctuations 
in magnetization between different gauge configurations, adding less than one part in $10^4$ to the susceptibility, which
is completely negligible in comparison to the ordinary statistical errors in this study.  
Thus, switching from PBC to OBC would seem to transform use of Coulomb gauge with Monte
Carlo data from a relatively rough and problematic method into a precise one, where the additional spurious fluctuations 
in quantities
due to imprecision in gauge fixing are negligible compared to the ordinary thermal (quantum) fluctuations.

Another test used for finding systematic error was to vary the Metropolis hit size for the run, using the closely-coupled
algorithm where the gauge relaxation is performed after each sweep.  This is the maximum amount by which 
new gauge elements are allowed to differ from old ones in an update.  Hit sizes vary from -1 to 1, with -1 meaning
no change allowed at all and 1 meaning the full group (it is the negative of the minimum $a_0$ component 
of the update matrix). One
would expect that the gauge setting algorithm would perform better when smaller changes were being made, since
it would be less likely to find a local minimum far from the global one, and basically have more tries on a more
restricted set of configurations.  These runs were compared with those
using a single relaxation after a random gauge transformation, 
and also the 
best of five and best of ten relaxations after random gauge transformations.  The latter were only run at 
a hit size of zero because
they would not be expected to show a dependence on hit size. The results are shown in Fig.~3 for 
Binder cumulant and magnetization.  It is clear that
there is very little difference between any of these algorithms - even the single relaxation after random gauge 
transformation gives results consistent with the others.  For
the magnetization, a slightly larger value is obtained with best of ten method than for a single relaxation. The best of ten
value agrees with that from
the closely-coupled
algorithm.  This suggests that the closely-coupled algorithm with zero hit size, as used in this paper is competitive
with the best of ten algorithm. However the differences are so small it appears that even the single relaxation after random
gauge transformation could be used in the OBC case. These runs were long enough (50,000 sweeps) that 
they should have  encountered
exceptional 
configurations if such still exist, one or two bad values from which would have had a large effect on the higher moments.
So the data seem to indicate that the previous
difficulties with Coulomb gauge relaxations no longer occur when using open boundary conditions. Indeed using open boundary 
conditions opens the path to highly precise measurements in the Coulomb gauge in which the efficacy of the gauge fixing 
algorithm is no longer a factor.
\begin{figure}
                      \includegraphics[width=2.5in]{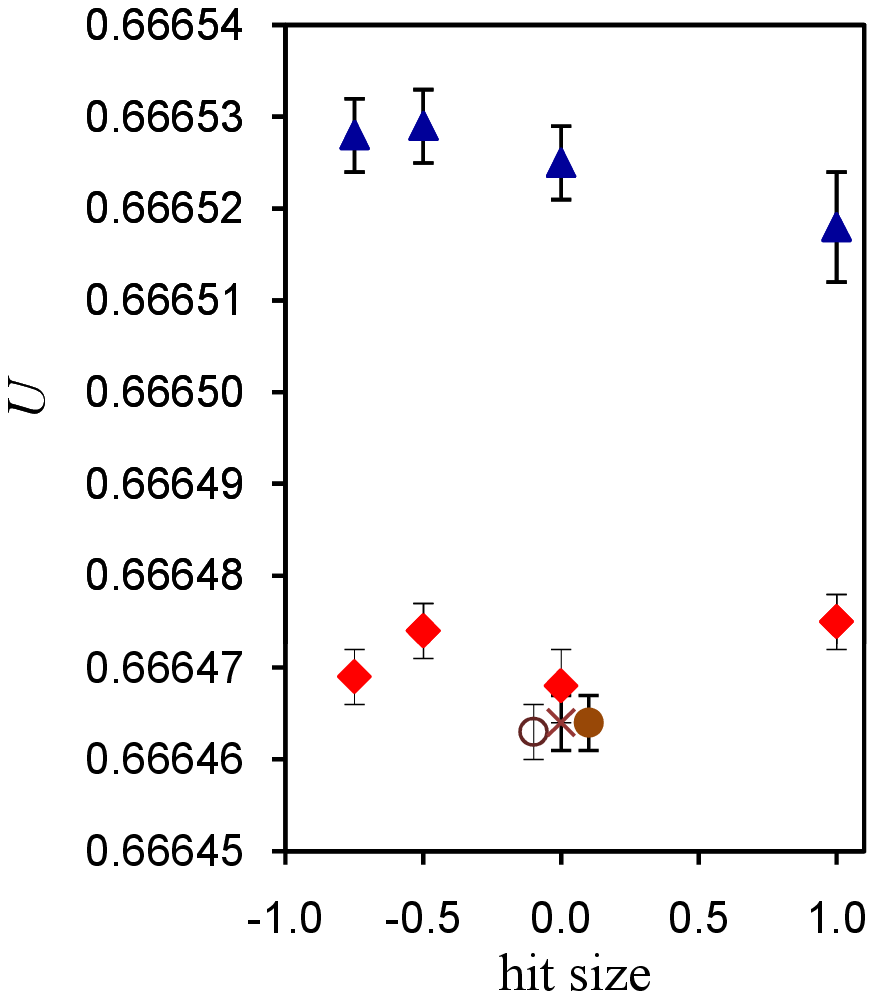}\includegraphics[width=2.5in]{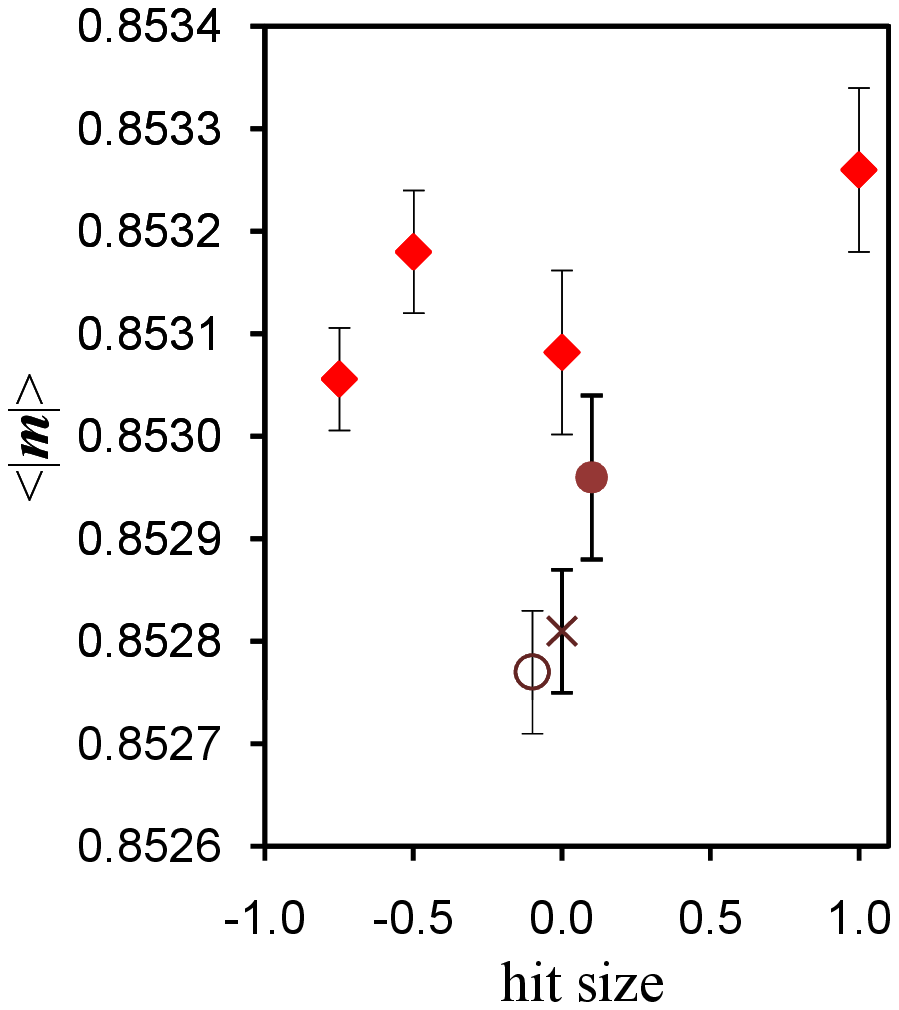}
                                 \caption{Binder cumulant, $U$, and magnetization vs. hit size for the closely-coupled
algorithm on $16^4$ lattices at $\beta =3.5$ (diamonds). Also shown is relaxation after a single random 
gauge transformation (open circle), best of five 
($\times$) and best of 10 (filled circle).  Points are horizontally offset slightly at hit size 0 for clarity. Also shown
are $24^4$ values for $U$ (triangles), demonstrating that algorithmic differences, if they exist, are much smaller than
the gap between $16^4$ and $24^4$ values.}
          \label{fig3}
       \end{figure}

\section{Expanding the parameter space to connect the spin model to the gauge theory}

The possible existence of a zero-temperature phase transition in the usual 4-d SU(2) gauge theory
can be put into clearer focus by considering a larger coupling space in which horizontal and vertical 
plaquettes have different coupling parameters, $\beta _H$, $\beta _V$.  Vertical plaquettes are those that include 
the fourth (Euclidean time) direction. As discussed above, for $\beta _H = \infty$ horizontal links are locked
to unity in the Coulomb gauge, which transforms the 4-link gauge interaction into a 2-link nearest-neighbor 
spin interaction (scalar product) among the vertical links.  Thus the theory is exactly a set of multiple 
non-interacting copies of the O(4) 3-dimensional Heisenberg model  (the quantum version of which is the 
non-linear sigma model). A similar connection of the SU(2) gauge theory to the Heisenberg spin model in one fewer dimensions
was explored analytically in \cite{frohlich}.  The 3-d O(4) classical Heisenberg model has a well-known higher-order
ferromagnetic phase transition at $\beta _V = 0.936(1)$ with critical exponents $\nu = 0.7479(90)$, 
$\gamma /\nu = 1.9746(38)$, $\beta / \nu = 0.5129(11)$,
and $\alpha = 2-d \nu = -0.244(27)$ \cite{O4}. Since $\alpha <0$ there is no infinite singularity in the specific heat, but rather a cusp. 
As mentioned before, the long-range ferromagnetic order of the Heisenberg model is quite robust to the addition
of a significant number of randomly chosen antiferromagnetic links, deleted links, or added next 
neighbor interactions\cite{robust,diluted}.  (Although the O(3) model is usually studied, the O(4) model appears to differ
only in detail.)  Thus, it would be quite surprising if the transition did not enter the ($\beta _V$,$\beta _H$)
coupling plane as $\beta _ H$ is backed off from $\infty$. In other words one would expect the transition to still exist
for large but finite $\beta _H$.  In this region, the horizontal gauge fields will still be very close to unity, so the 
Hamiltonian would still be primarily the Heisenberg nearest-neighbor interaction. Small additional terms of order
$1/\beta _H$ would be present which would add disorder and likely add next-neighbor interactions if the horizontal
gauge fields were to be integrated out. A renormalization of the pseudotemperature ($1/\beta _V$) would also 
be expected. Because the integrated horizontal
gauge fields do not break the remnant O(4) symmetry which remains after Coulomb gauge fixing, these effective 
spin interactions cannot 
favor domain formation as a random external magnetic field would\cite{rmf}.  If the phase transition does enter
the coupling plane for $\beta _H < \infty$, then due to its symmetry-breaking nature it cannot end except at another
edge of the phase-plane (it must divide the phase-plane into two disconnected regions of different symmetry). As $\beta _H$ is
lowered from infinity, the critical value of $\beta _V$ must increase to compensate for the additional disorder, so clearly
these must then meet at a finite value $\beta _c = \beta _{Vc} = \beta _H$, which would be the critical point
of the SU(2) gauge theory.

Since a ferromagnetic phase in Coulomb gauge is necessarily deconfined\cite{zw}, this calls into question the customary assumption
of confinement in the continuum limit ($\beta \rightarrow \infty$). If the conventional interpretation of no 
phase transition in SU(2) is correct, then the O(4) critical
point would have to exist only for $\beta _H = \infty$ and cease to exist for any $\beta _H < \infty$. 
In other words the slightest
$ 1/\beta _H >0$ coupling must destroy the long-range order. Since the large $\beta _H$ case turns out to be
simpler and more easily studied
than the $\beta _V = \beta _H$ symmetric case, it seems worth focusing on it first to investigate this important question.
It might, for instance, be possible to extend the proof of long range order in the Heisenberg model\cite{prooflro} 
to the small
$1/\beta _H$ region, which by the above argument would prove the existence of an SU(2) transition.

A Monte Carlo simulation at $\beta _H = 20$ was performed to look for a Heisenberg-like transition in the large 
$\beta _H$ region. Such a transition is easily found using the Coulomb gauge method with 
open boundary conditions.  As previously
mentioned, the order parameter is simply the fourth-direction pointing links which act as O(4) spins. The ``gold standard"
method of finding a Binder cumulant crossing from simulations on different size lattices is 
used to find the position 
of the phase transition on the infinite lattice, just as is done for the case of the spin model itself.  The Binder
cumulant\cite{binder}, $U = 1-<|\vec{m}|^4>/(3<|\vec{m}|^2 >^2 )$, is plotted in Fig.~4, showing a definite crossing near 
$\beta _V =1.01$.
\begin{figure}\centering
                      \includegraphics[width=3.5in]{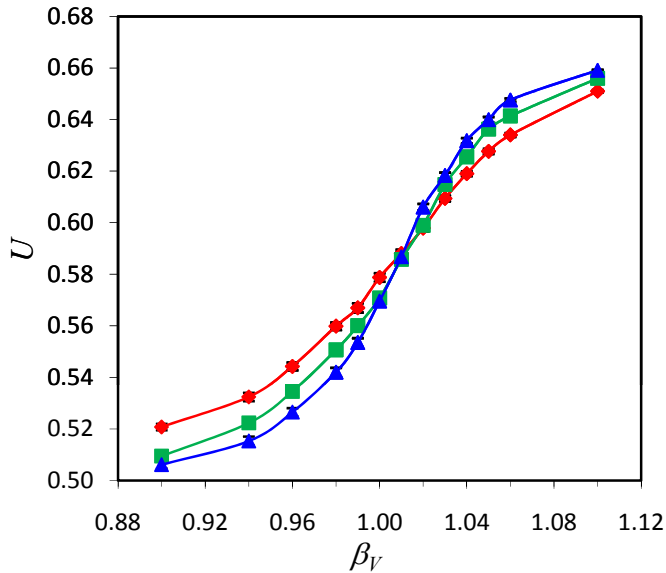}
                                 \caption{Binder cumulant crossing for the $\beta _H =20$ theory. $U_{24}$ exceeds $U_{20}$
by $2.5\sigma $ and $U_{16}$ by $9\sigma $ at $\beta = 1.06$.  Diamonds are $16^4$, squares $20^4$, and triangles $24^4$ 
lattices on this and all subsequent plots.}
          \label{fig4}
       \end{figure}

For the O(4) order parameter, Binder cumulant values range from 
0.5 in the deep paramagnetic region to 2/3 in the deep ferromagnetic region\cite{mdop}.  Fig.~5 shows scaling plots
for Binder cumulant, $U$, subtracted susceptibility $\chi =V(<|\vec{m}|^ 2>-<|\vec{m}|>^2 )$, and 
magnetization $<|\vec{m}|>$,
demonstrating good collapse to finite size
scaling ans\"{a}tze \cite{fss} (see axis labels for detailed scaling functions, 
with $T= 1/\beta _V$ and $T_c = 1/\beta _{Vc}$).
\begin{figure}
                      \includegraphics[width=2.5in]{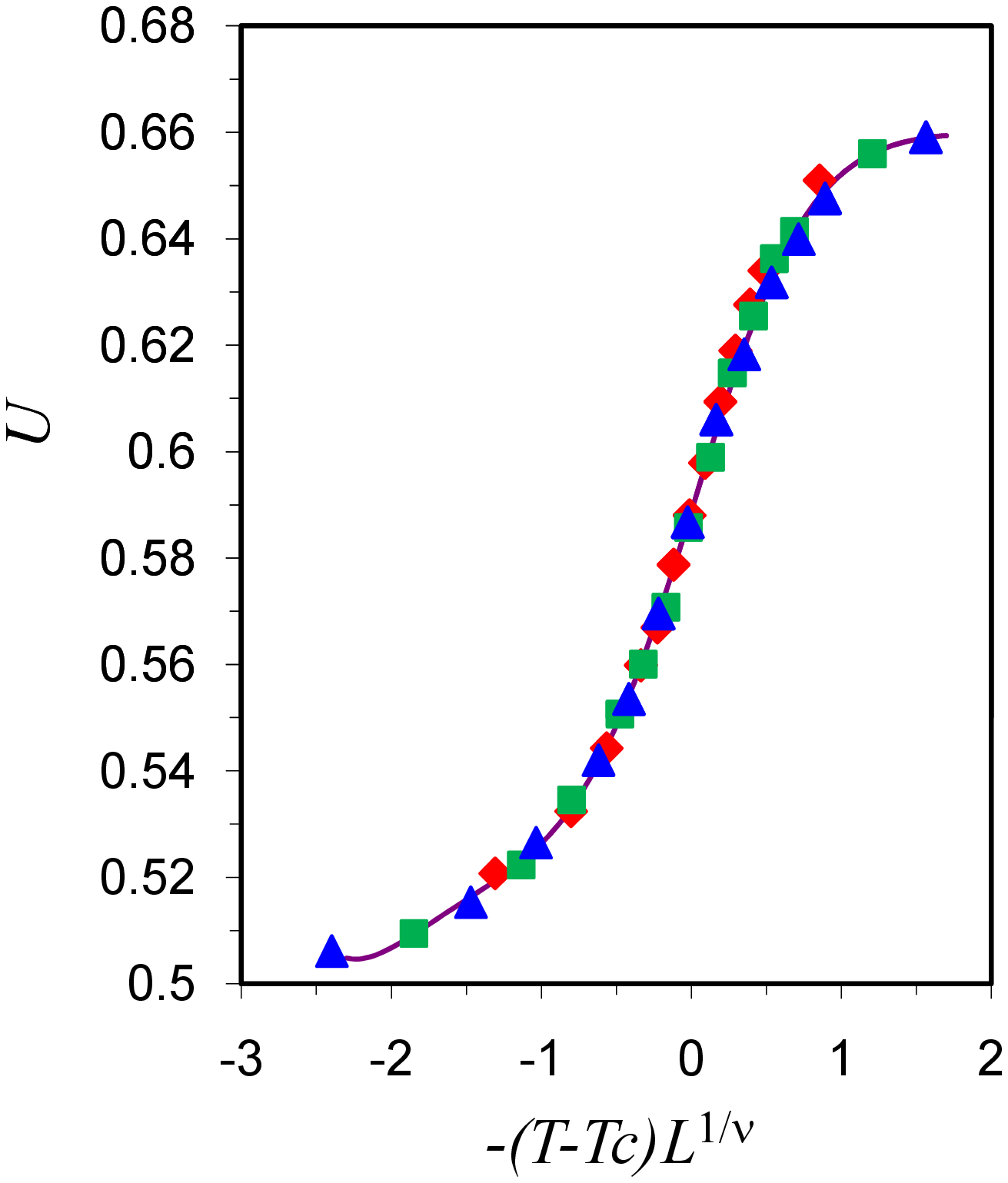}\includegraphics[width=2.5in]{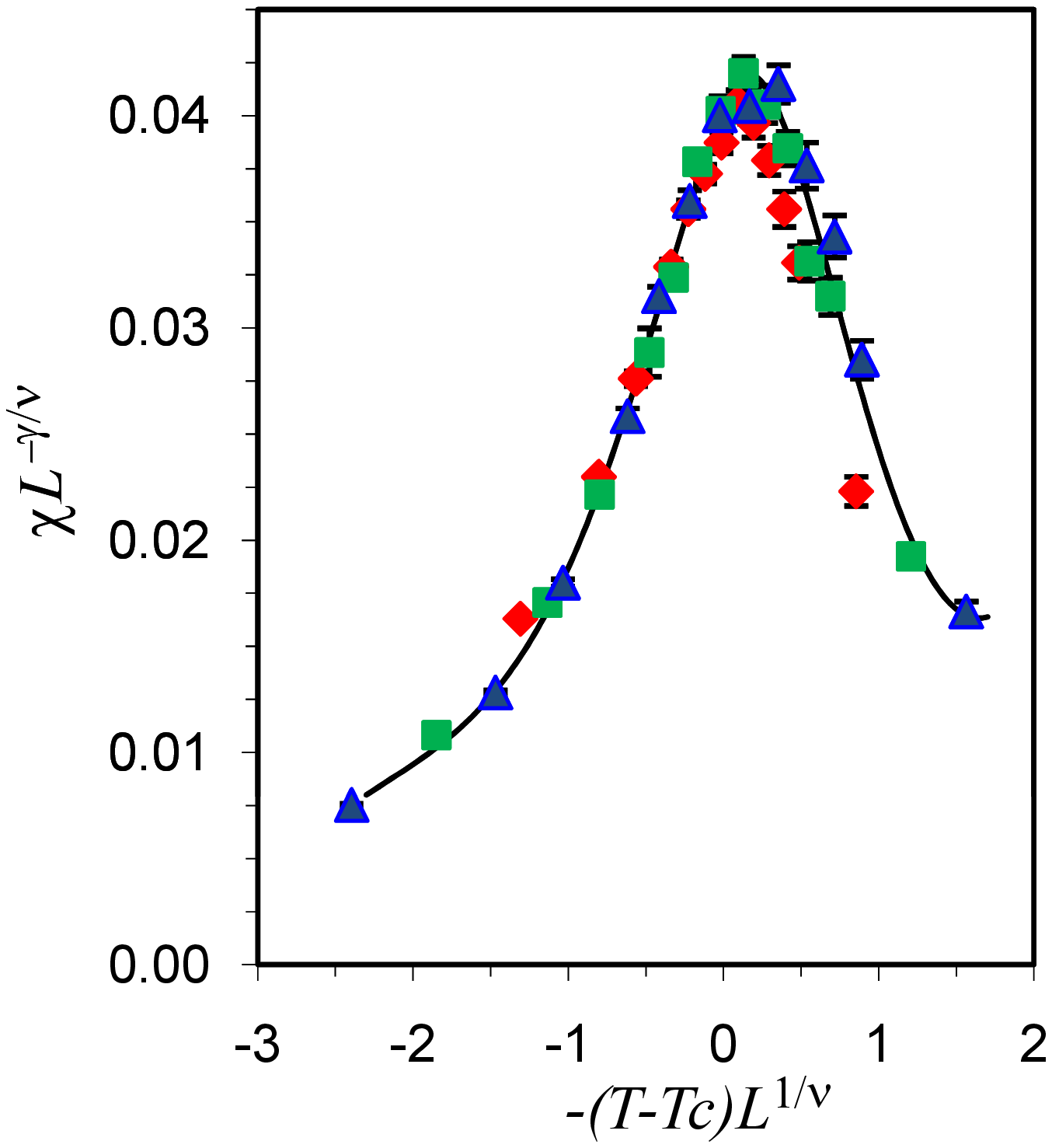}
\includegraphics[width=2.5in]{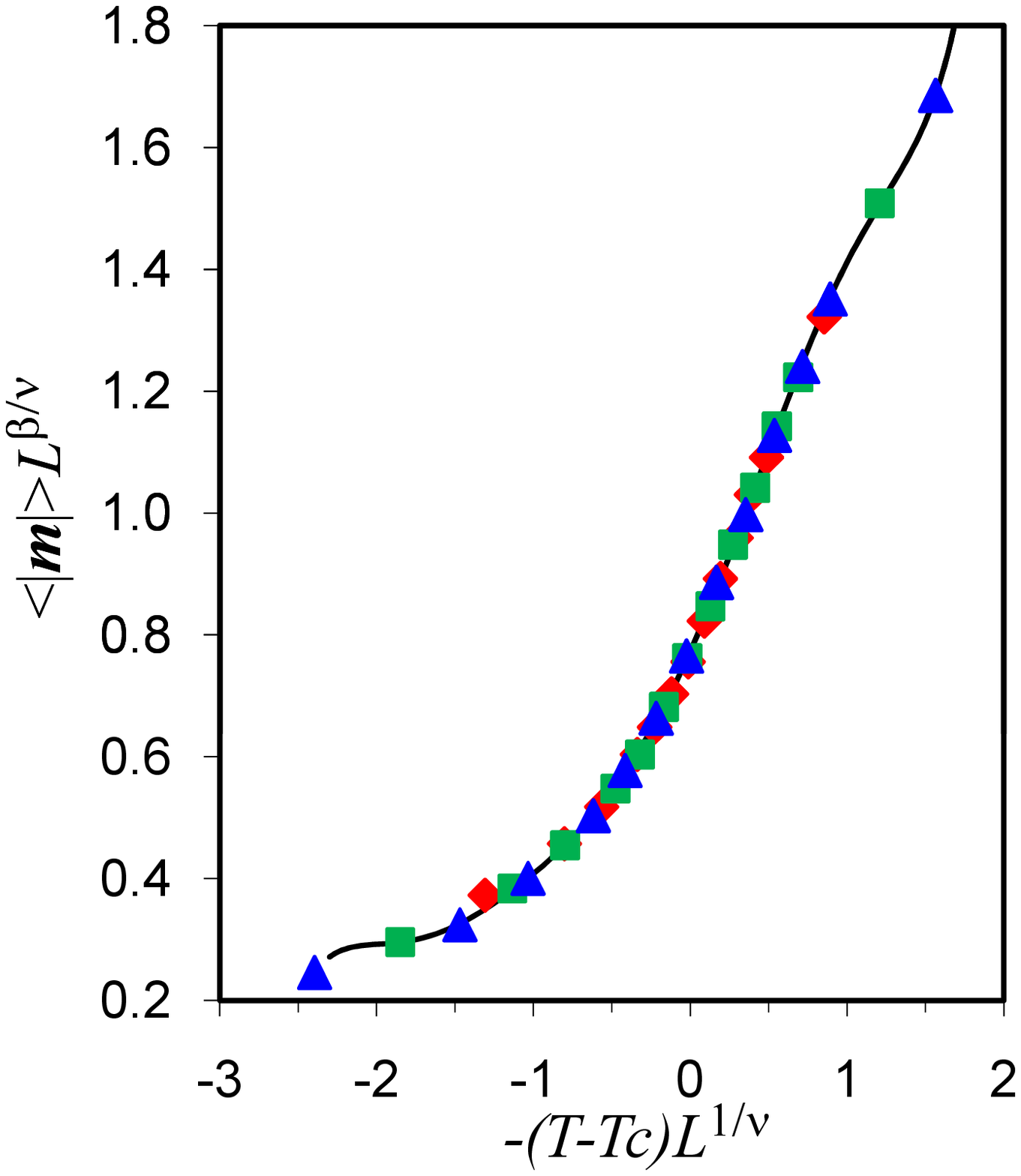}\includegraphics[width=2.5in]{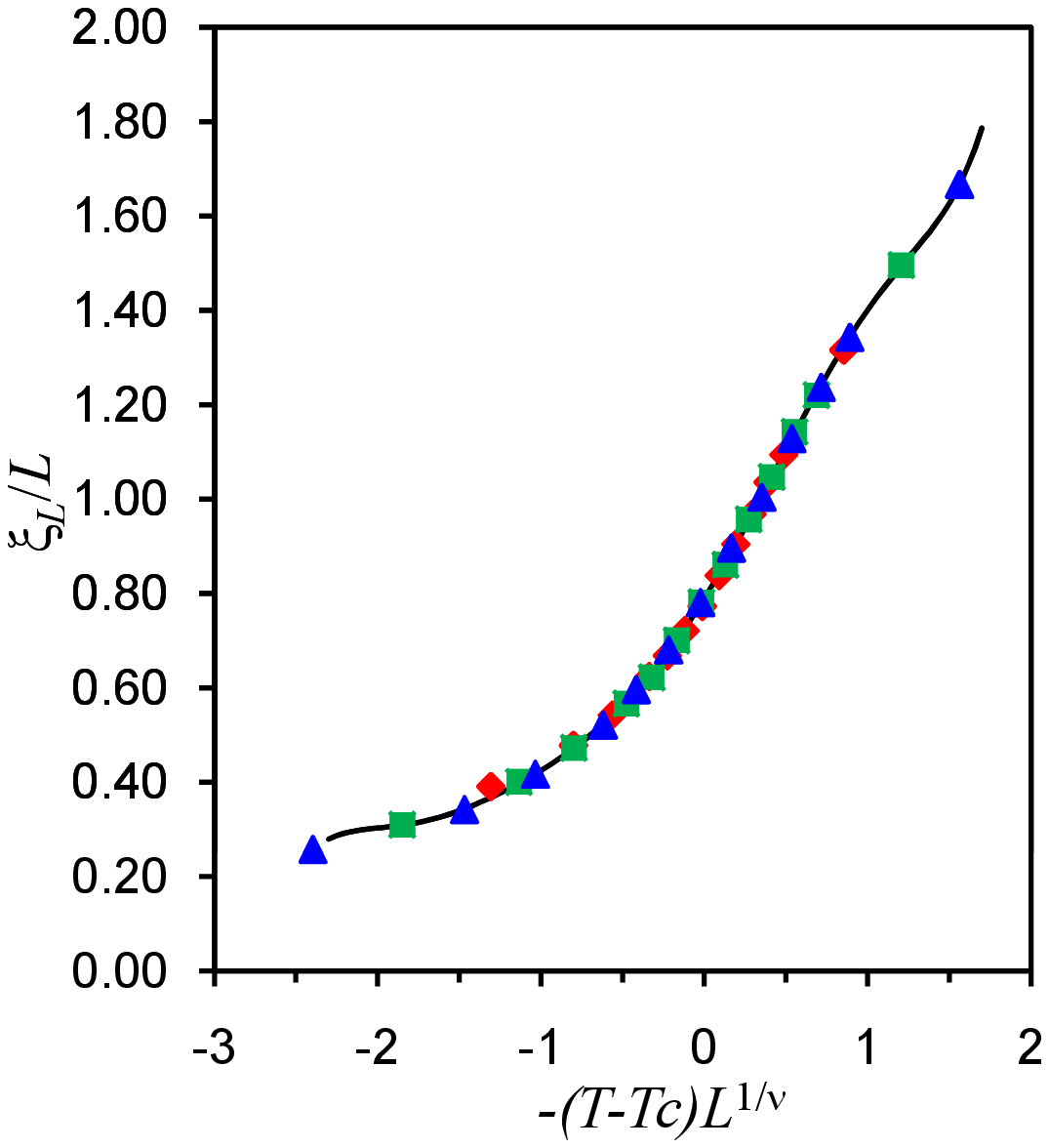}
                                 \caption{Finite size scaling collapse plots for $U$, $\chi$, $<|\vec{m}|>$,
and $\xi _L /L$ for the $\beta _H =20$ theory.}
          \label{fig5}
       \end{figure}
  
One disadvantage of using open boundary conditions
is that lack of translational invariance makes the Fourier transform impossible, and thus the second moment correlation 
length\cite{smcl}, which also yields scaling information, cannot be defined. However, the unsubtracted susceptibility,
$\chi _{b}= V<|\vec{m}|^2>$
when raised to the appropriate power can be used as a surrogate for the correlation length, 
\begin{equation}
\xi _L = \chi_b ^{\nu/\gamma} .
\end{equation}
The subscript $L$ refers to the linear lattice size.  
The quantity $\xi _L /L$ can be then used in the collapse fits to further constrain the determination of exponents $\gamma$
and $\nu$ (see Fig.~5d).  Once $\gamma /\nu $ has been determined, $\chi _L /L$ may be used to see crossings at a fixed point
corresponding to the infinite
lattice critical point.  Fig.~6 gives a very clear crossing at a coupling consistent with that of the Binder cumulant.  The overall
fit giving the best collapse of the four quantities, shown in Fig.~5, gives $\beta_{Vc} =1.01(2)$, 
$\nu = 0.97(17)$, $\gamma /\nu = 2.00(25)$,
$\beta / \nu = 0.51(14)$, with an overall $\chi ^2/d.f. = 2.4$.  Only the $20^4$ and $24^4$ data are used in the 
scaling fits as the $16^4$ appears a bit too small
to follow the universal scaling functions within the statistical errors here (higher order corrections are sometimes 
needed for smaller lattices).  
\begin{figure}\centering
                      \includegraphics[width=3.5in]{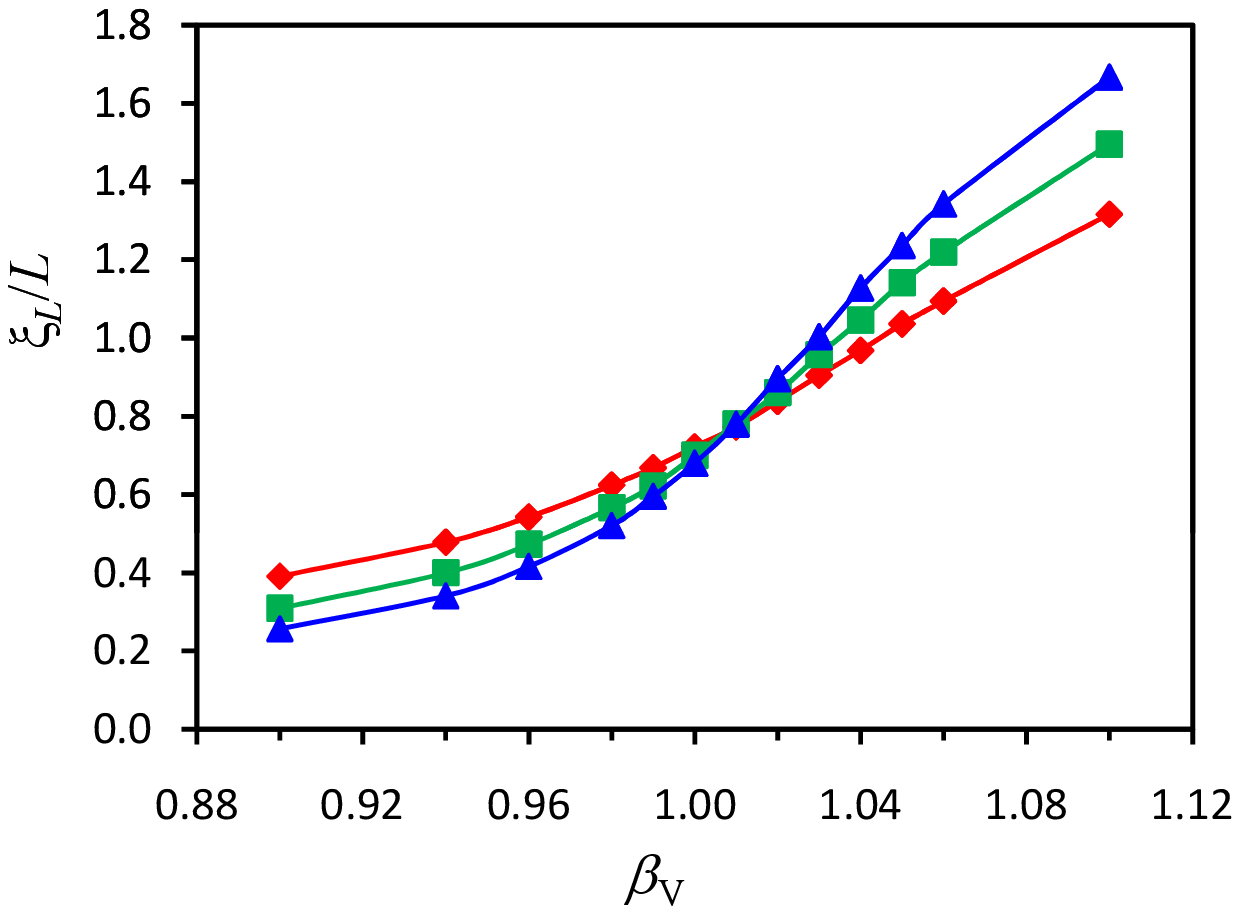}
                                 \caption{Crossing graph for $\xi _L /L$ for $\beta _H =20$. The $24^4$ value exceeds the
$20^4$ by $21\sigma $ and the $16^4$ by $43\sigma $ at $\beta = 1.06$.}
          \label{fig6}
       \end{figure}
These runs were all for 50,000 sweeps following 10,000 equilibration
sweeps. The $\gamma / \nu$ and $\beta / \nu$ values are consistent with those for the O(4) Heisenberg 
model, whereas the value of $\nu $ appears somewhat larger.  This makes sense considering that the value of $\nu $ seen
below for the $\beta _H = \beta _V $ SU(2) case is much higher, around 3.42.  Thus the $\beta _H = 20$ transition appears
to have already started evolving in this direction but is otherwise very similar to the Heisenberg transition itself.
Of course the main result here is that the $\beta _H = 20$ transition simply exists at a finite $\beta _{Vc}$. 
As argued above, due to the symmetry breaking nature of the phase transition, this is a sufficient condition for 
the existence of an SU(2)
transition at finite $\beta $.  Binder cumulant crossings and scaling collapse plots are tried and true methods for finding
phase transitions.  Calling into question the $\beta _H = 20$ result would also seem to call into question the Heisenberg 
results themselves, since the methods and data are so similar.  The reliability of these methods has been demonstrated
in a large
variety of models.  If something drastically different were to happen on very large lattices, it is unlikely that such clean
crossings would have been observed here.  In order for the transition to disappear, the behavior of $U$ and $\xi _L /L$
in the weak coupling region would have to switch from the observed increasing function with lattice size to a decreasing function. 
Such a change in behavior would be highly unusual, if not unprecedented.

\begin{figure}\centering
                      \includegraphics[width=3.5in]{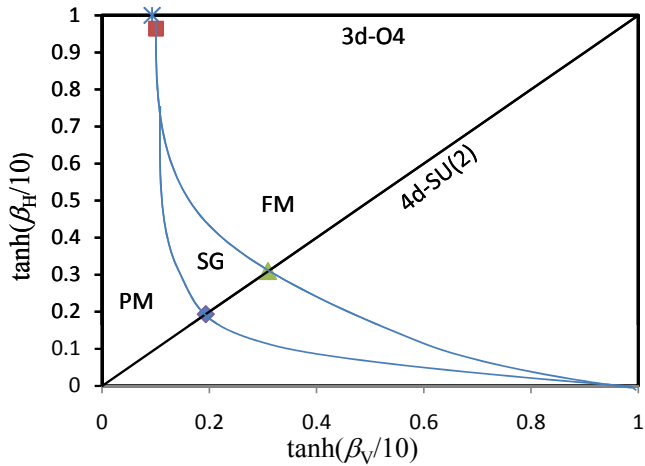}
                                 \caption{Suggested phase diagram from the phase transitions seen in this paper at
$\beta _H =20$ (square) and $\beta _H = \beta _V$ SU(2) gauge theory (triangle). Also shown are the 3-d O(4) Heisenberg
model transition on the upper axis (star) and the paramagnet to spin-glass transition seen in ref.\cite{su2sg} (diamond).}
          \label{fig7}
       \end{figure}
The suggested phase diagram in the extended plane is given in Fig.~7.  The Heisenberg transition is seen on the upper
boundary with the $\beta _H = 20$ transition, just described, below it.  Further down, along the $\beta _V = \beta _H $ 
symmetry line, is
the SU(2) transition at $\beta = 3.2$ to be described in the next section.  Also shown is the spin glass transition 
for SU(2)  at $\beta = 1.96$
described in \cite{su2sg}. The point where the spin-glass phase first splits off has not yet been determined, nor has the 
behavior below the symmetry line.  The bottom boundary is a set of one-dimensional O(4) spin models, which are 
in the unbroken
phase except at $\beta _V = \infty $.  So a likely place for the transitions to end up is this point, but  further
investigation is needed.

\section{Monte Carlo simulations}

A similar study was performed for the regular SU(2) theory, again using open boundary conditions.
Some additional details, which also apply to the simulations of the last section are given here.
Although opening the boundary conditions might seem a rather drastic step, when larger
lattices are used it is certainly practical and causes fewer difficulties than one might expect.  The outer three layers
were discarded in order to lessen the effects of the open boundary. The vast majority of the boundary-effect occurs here.
In other words, for a $16^4$ lattice quantities are only measured on the inner $10^4$ and for finite-size scaling
purposes the size is $10^4$.  This can be thought of as a soft open boundary.
All  remaining boundary effects can be absorbed into the finite size scaling applied to different lattice sizes. In
principle one does not need to exclude boundary layers, but that would require larger lattices for sensible results.
One also
loses the translational invariance, which, as mentioned above, prevents the 
calculation of the second moment correlation length.
Otherwise analysis is relatively unchanged.

Lattices of $16^4$, $20^4$ and $24^4$ were measured for 50,000 sweeps, after an initial equilibration of
10,000 sweeps, for $\beta$ between 2.3 and 3.6 . The gauge overrelaxation was very slow at $\beta = 2.3$,
taking roughly 4000 sweeps per Monte Carlo sweep.  For $\beta > 3.0$, 100 or fewer overrelaxation sweeps
were needed. Overrelaxation was terminated when the per-link value of the gauge condition
changed by less than $2\times 10^{-8}$. Test runs with a criterion ten times smaller showed no difference.
Of course for many of these large $\beta $ the correlation length is
larger than the largest lattice size. Simulations designed to explore continuum physics must be run
in a region where the correlation length is smaller than the lattice to avoid finite lattice and finite temperature
effects.  However, what is being sought here is a possible critical point, where the correlation length is
infinite. So, just as in the study of magnetic theories, one must march right through the suspected critical
region with finite lattices and interpret the data using finite-size scaling.  If the true critical point is at
$\beta = \infty$ then finite-size scaling should be able to see that too. In that case, the scaling in the entire
range studied should look like the unbroken region with no apparent critical point. In particular, the Binder
cumulant should everywhere be a decreasing function of lattice size.

\begin{figure}\centering
                      \includegraphics[width=4in]{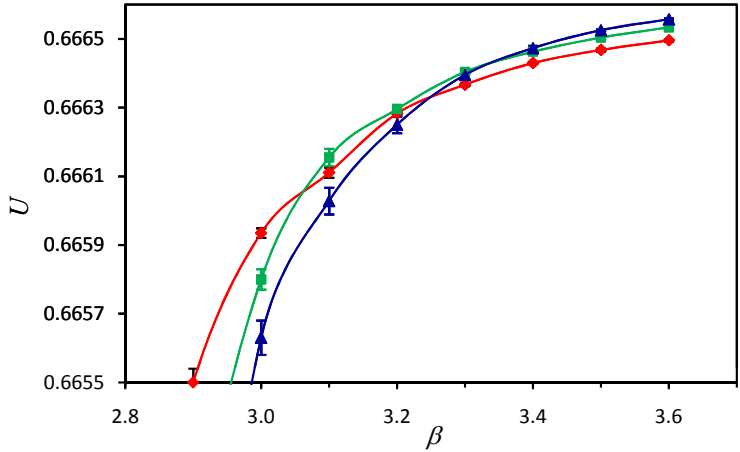}
                                 \caption{Binder cumulant crossing for the 4-d SU(2) theory.}
          \label{fig8}
       \end{figure}
\begin{figure}\centering
                      \includegraphics[width=4in]{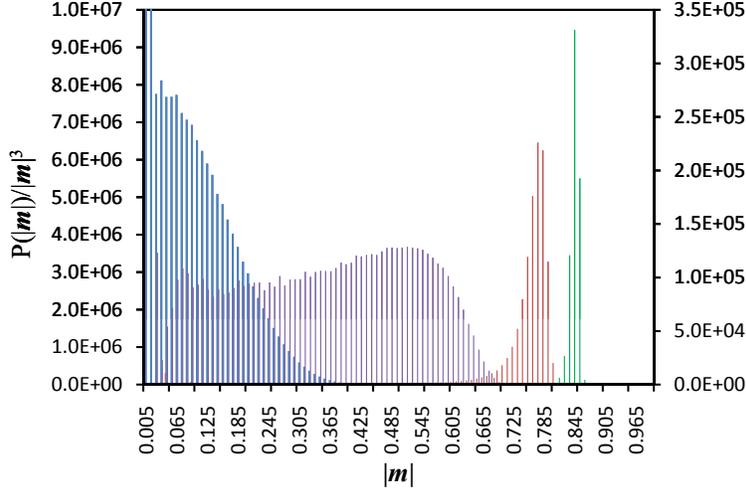}
                                 \caption{Magnetization histograms at $\beta = 2.3$, 2.5, 2.9 and 3.5 on $24^4$
lattices. The latter three use the right-side vertical scale.}
          \label{fig9}
       \end{figure}
The Binder cumulant, $U$, shows an apparent crossing at quite a high cumulant value somewhere between $\beta $ = 3.1 and 3.3
 (fig.~8). 
The crossing  around $\beta = 3.2$ 
is a bit messy, but for $\beta = 3.4$ to 3.6 $U$ is definitely an increasing function of lattice
size indicating this is in the spontaneously broken region. Errors at these $\beta $'s are very small.
At $\beta = 3.4$ $U_{24}$ exceeds $U_{20}$ by 4$\sigma $ and $U_{16} $ by 11$\sigma $. At $\beta = 3.5$ these 
are $5\sigma $
and $14\sigma $ respectively.
Magnetization histograms are given in Fig.~9, showing
a pattern typical of a higher order transition.  
For this O(4) order parameter one must take 
into account the geometrical factor 
(from solid angle) that biases the distribution toward
larger magnitudes. In the unbroken phase, 
the distribution of magnetization moduli, $|\vec{m}|$, is
expected to be a factor of $|\vec{m}|^3$ times a 
Gaussian, $\exp (-|\vec{m}|^2/2 \sigma _{m} ^2)$. To more easily
see the Gaussian behavior, the probability distribution $P(|\vec{m}|)$ is 
obtained by histogramming, and
the quantity $P(|\vec{m}|)/|\vec{m}|^3$ is plotted.
The value of $|\vec{m}|$ for each 
bin is not taken at the center,
but at a value that would produce a flat 
histogram in an $|\vec{m}|^3$ distribution, regardless
of bin-size choice. This is 
\begin{equation}
|\vec{m}|^{3}_{\rm{bin}}=\frac{1}{4}\frac{(m_{2}^{4}-m_{1}^{4})}{(m_2 - m_1 )} .
\end{equation}
where $m_2$ and $m_1$ are the bin edges. This 
detail affects only the first couple of bins in the 
histograms.
Magnetization and subtracted susceptibility curves are shown in
Fig.~10 with the susceptibility showing a peak growing with lattice size and shifting toward weaker couplings. 
Scaling collapse fits were then performed
for $\chi$ and $<|\vec{m}|>$ using $U$ itself as the scaling variable.  This can be done since $U$ has no 
anomalous dimension, and
has the advantage of being a single parameter fit (no fitting of $\beta _c$ is needed). From these fits, values of 
$\gamma / \nu = 2.951(3)$ and $\beta / \nu = 0.0250(8)$ were found.  
Since these are consistent with the hyperscaling relationship
$\gamma / \nu + 2 \beta /\nu = d = 3$, it was decided to redo the fits with only a single degree of freedom by enforcing
hyperscaling(Fig. 11), which yielded $\gamma / \nu = 2.950(2)$ with $\chi ^2/\rm{d.f.} = 1.7$. 
\begin{figure}
                      \includegraphics[width=2.5in]{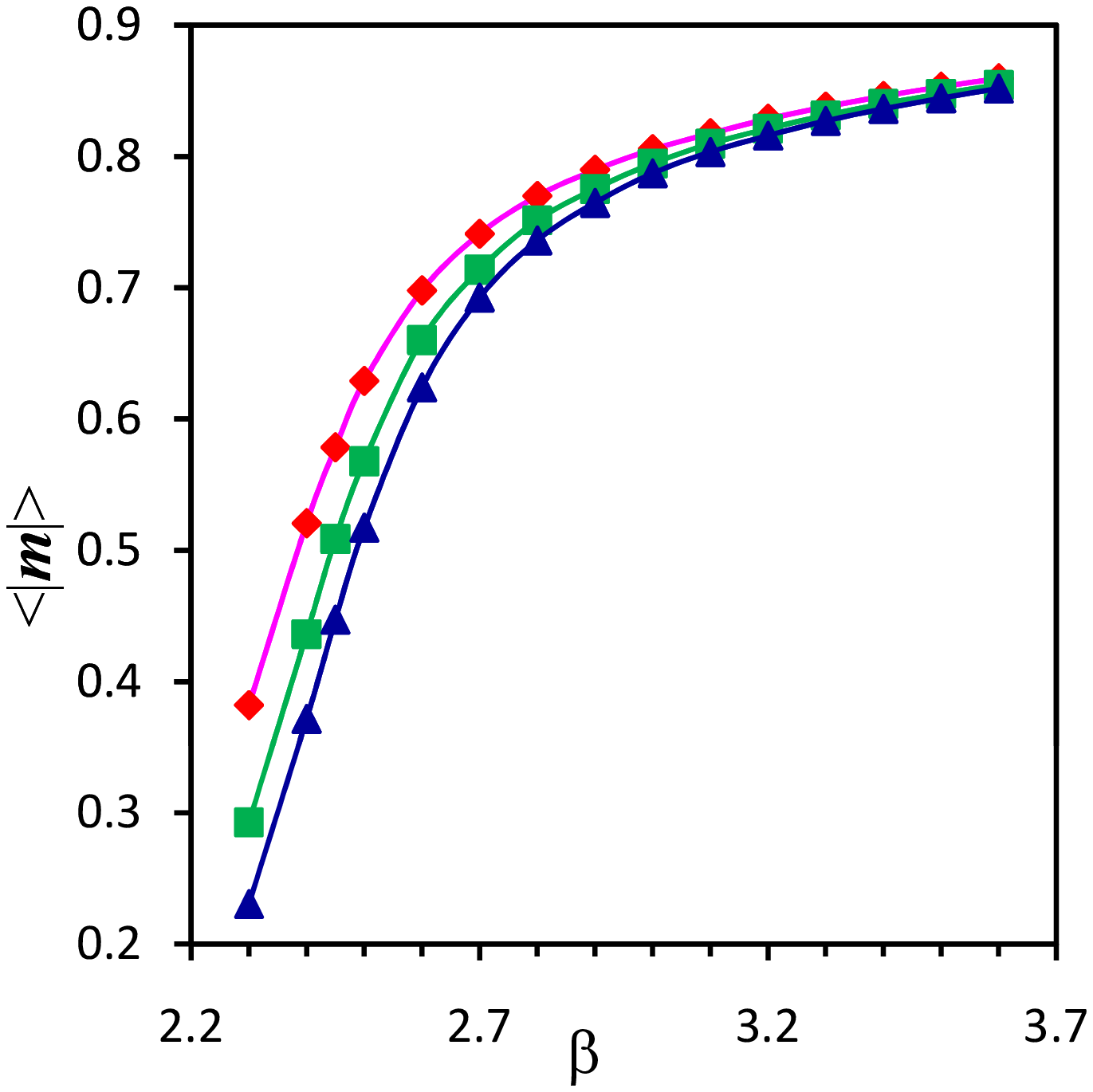}\includegraphics[width=2.5in]{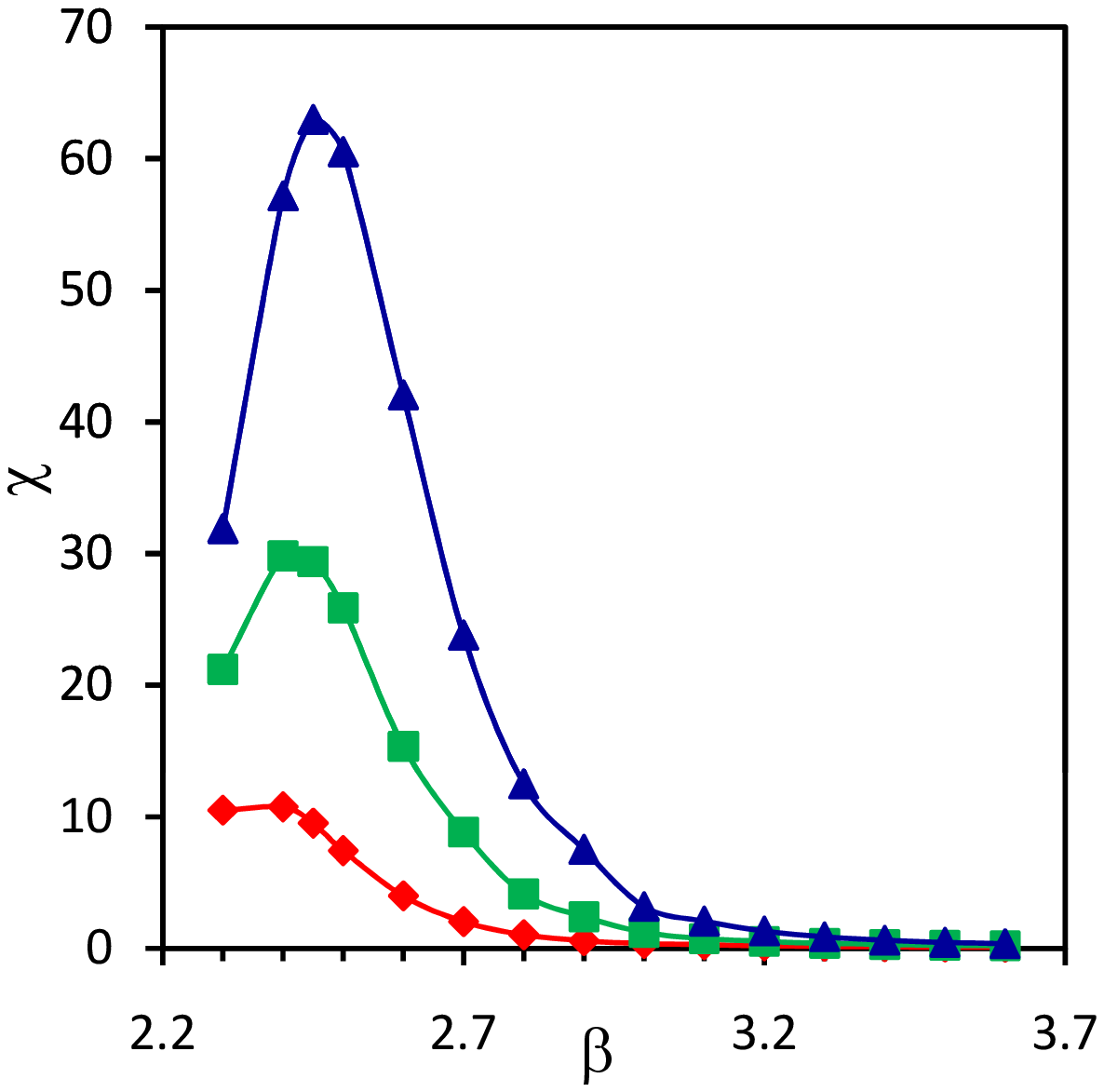}
                                 \caption{Magnetization and susceptibility for 4-d SU(2).}
          \label{fig10}
       \end{figure}
\begin{figure}
                      \includegraphics[width=2.5in]{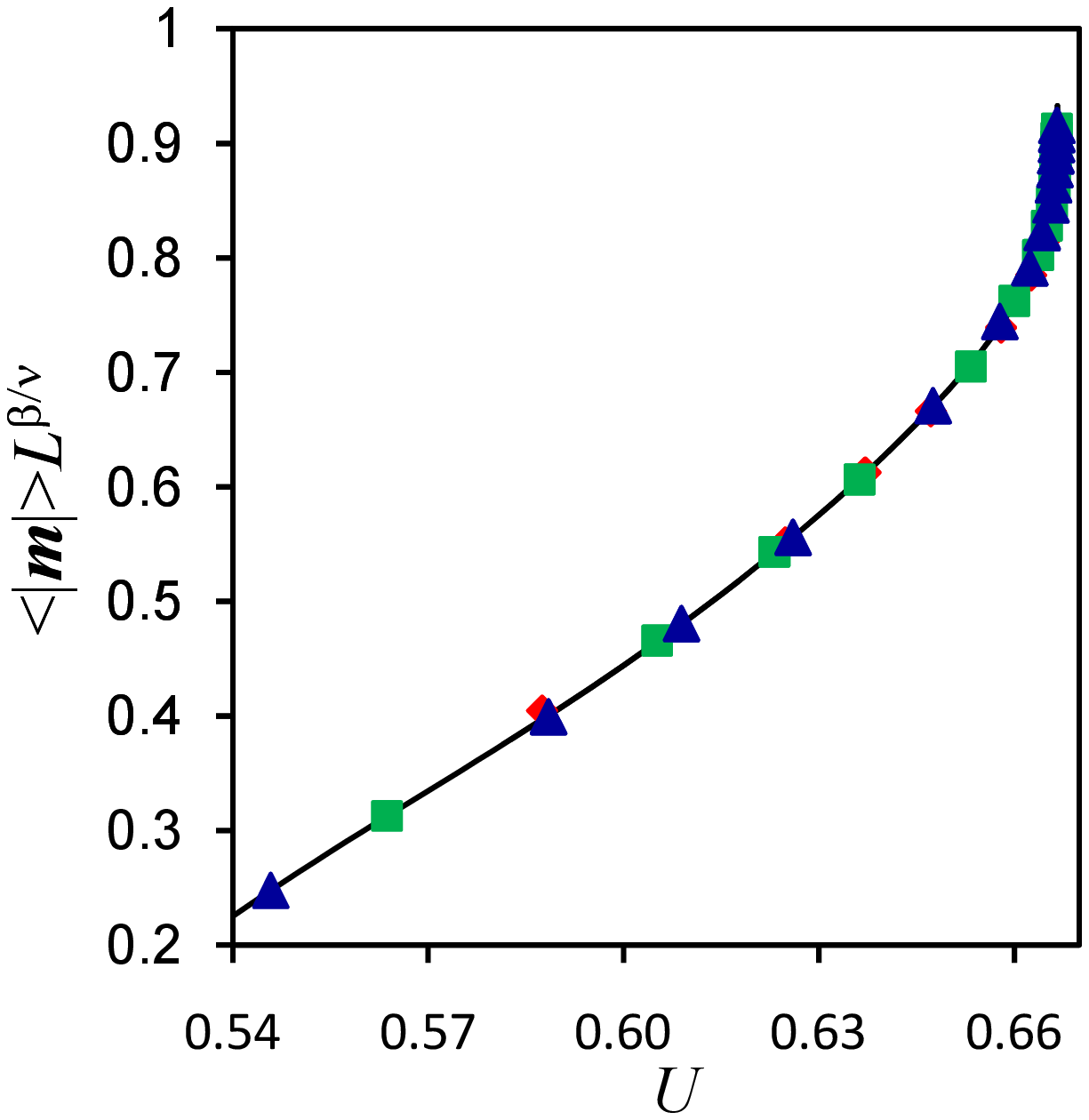}\includegraphics[width=2.5in]{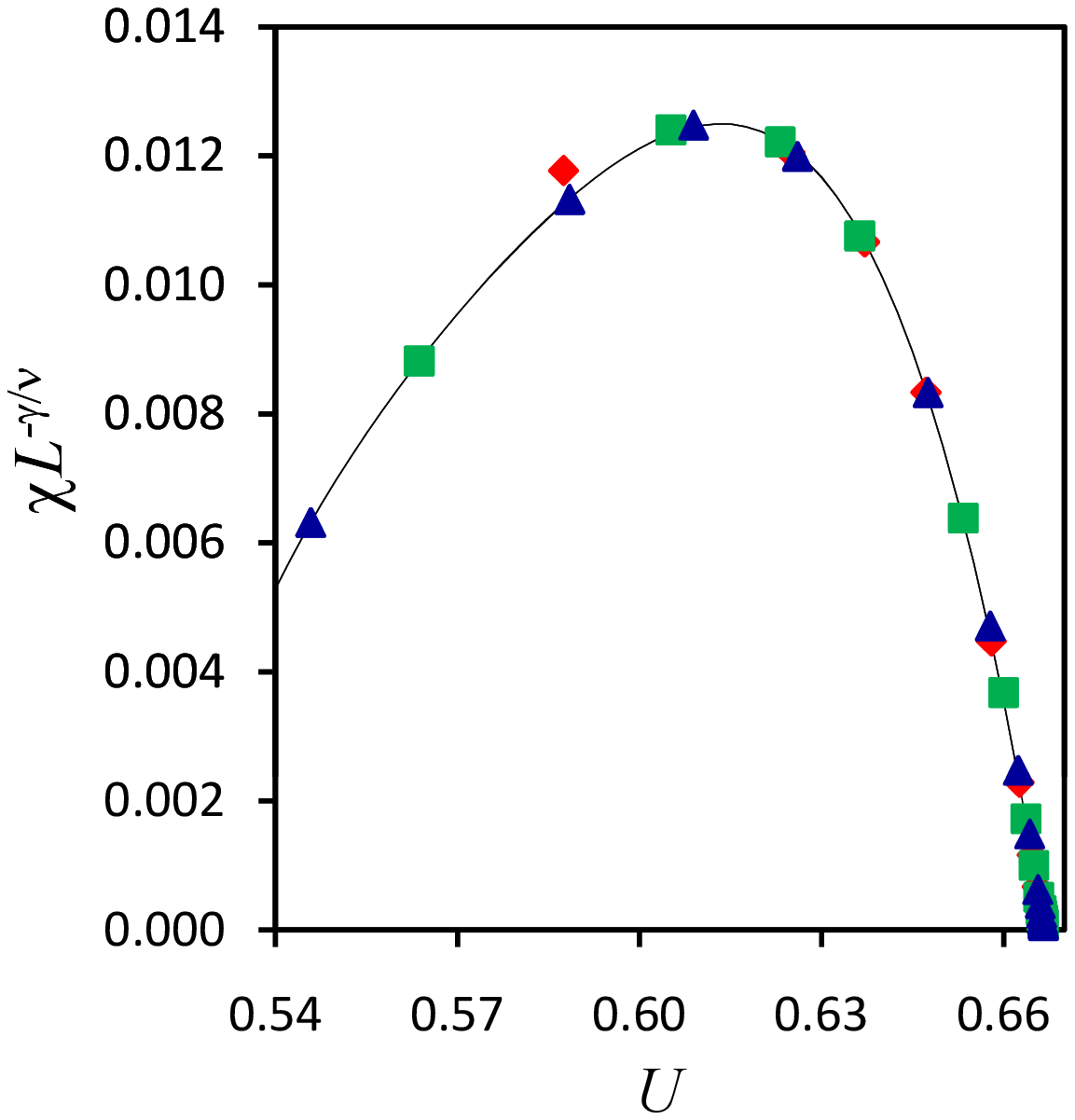}
                                 \caption{Magnetization and susceptibility scaling collapse fits using $U$ as
the scaling variable.}
          \label{fig11}
       \end{figure}
\begin{figure}\centering
                      \includegraphics[width=4in]{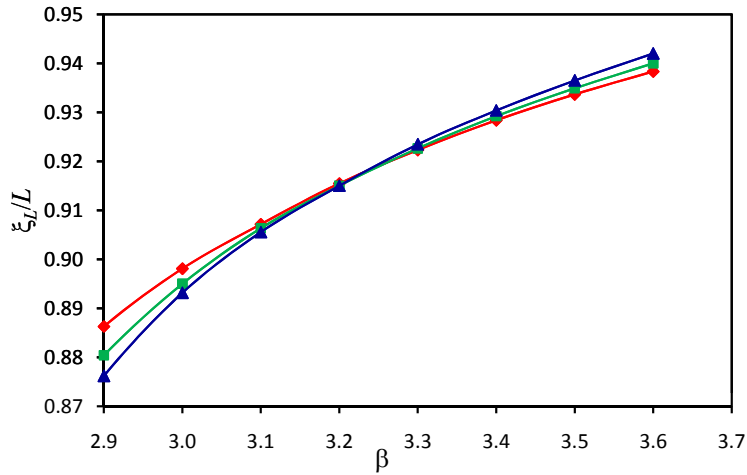}
                                 \caption{Crossing plot for $\xi _L /L$ for 4-d SU(2).}
          \label{fig12}
       \end{figure}
\begin{figure}
                      \includegraphics[width=2.5in]{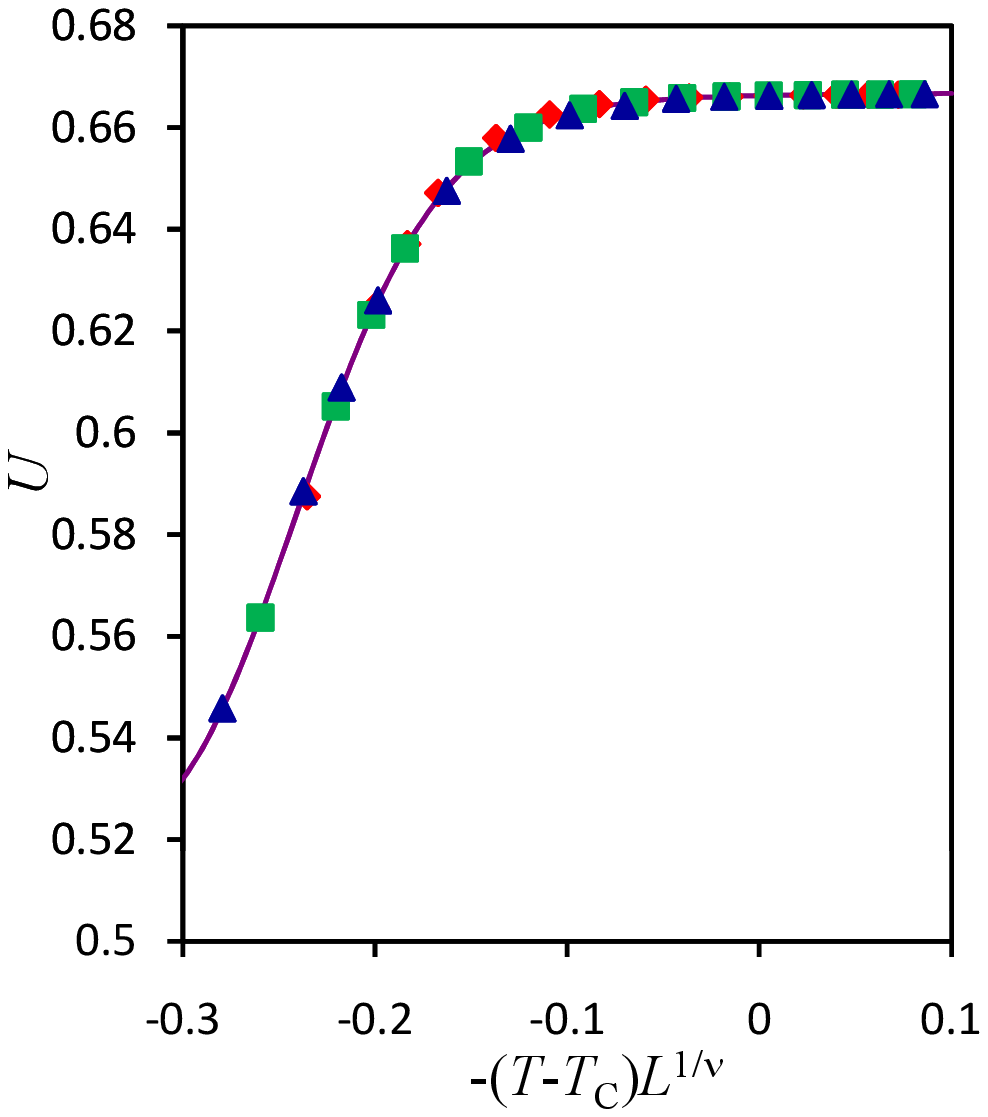}\includegraphics[width=2.5in]{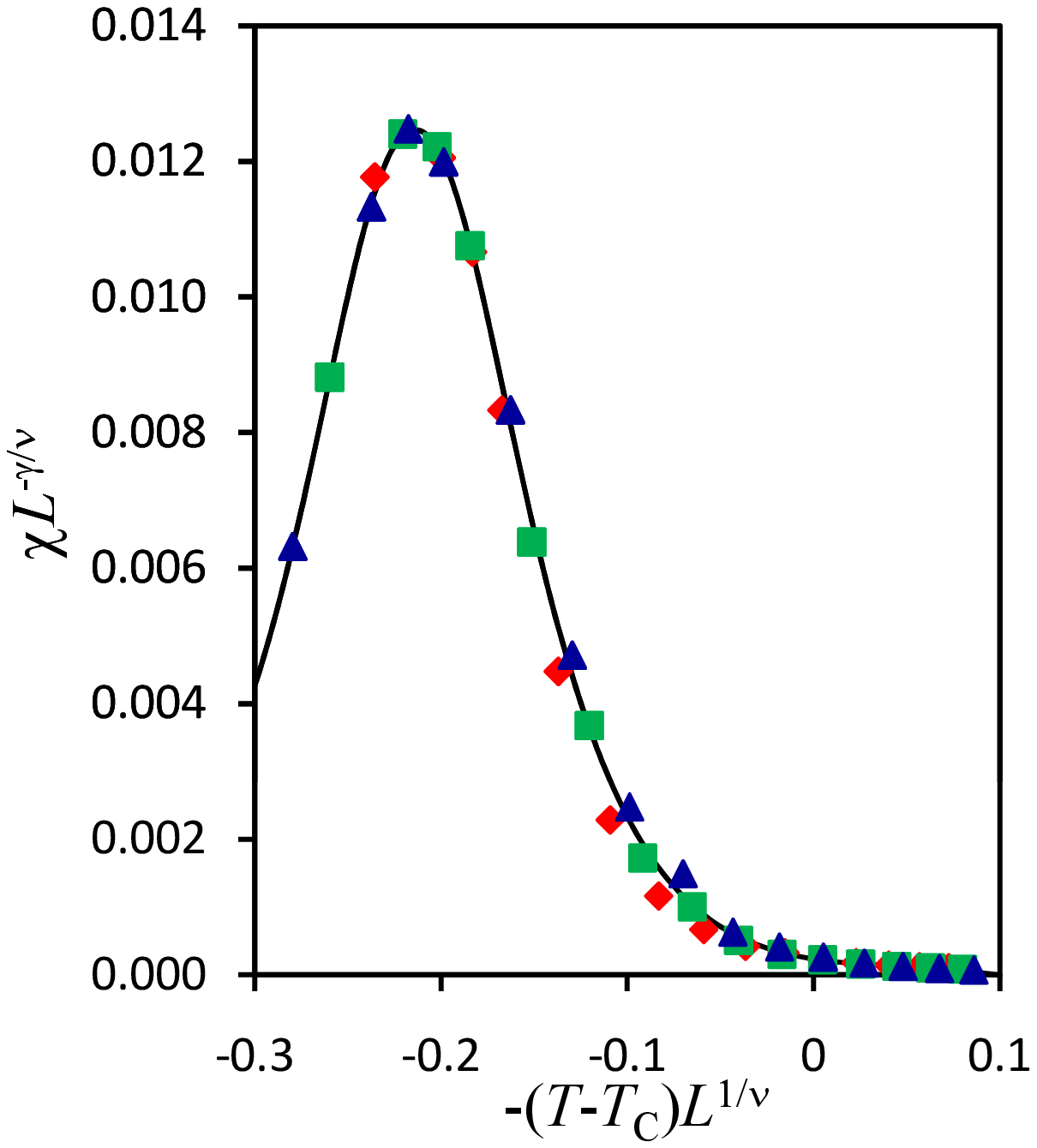}
\includegraphics[width=2.5in]{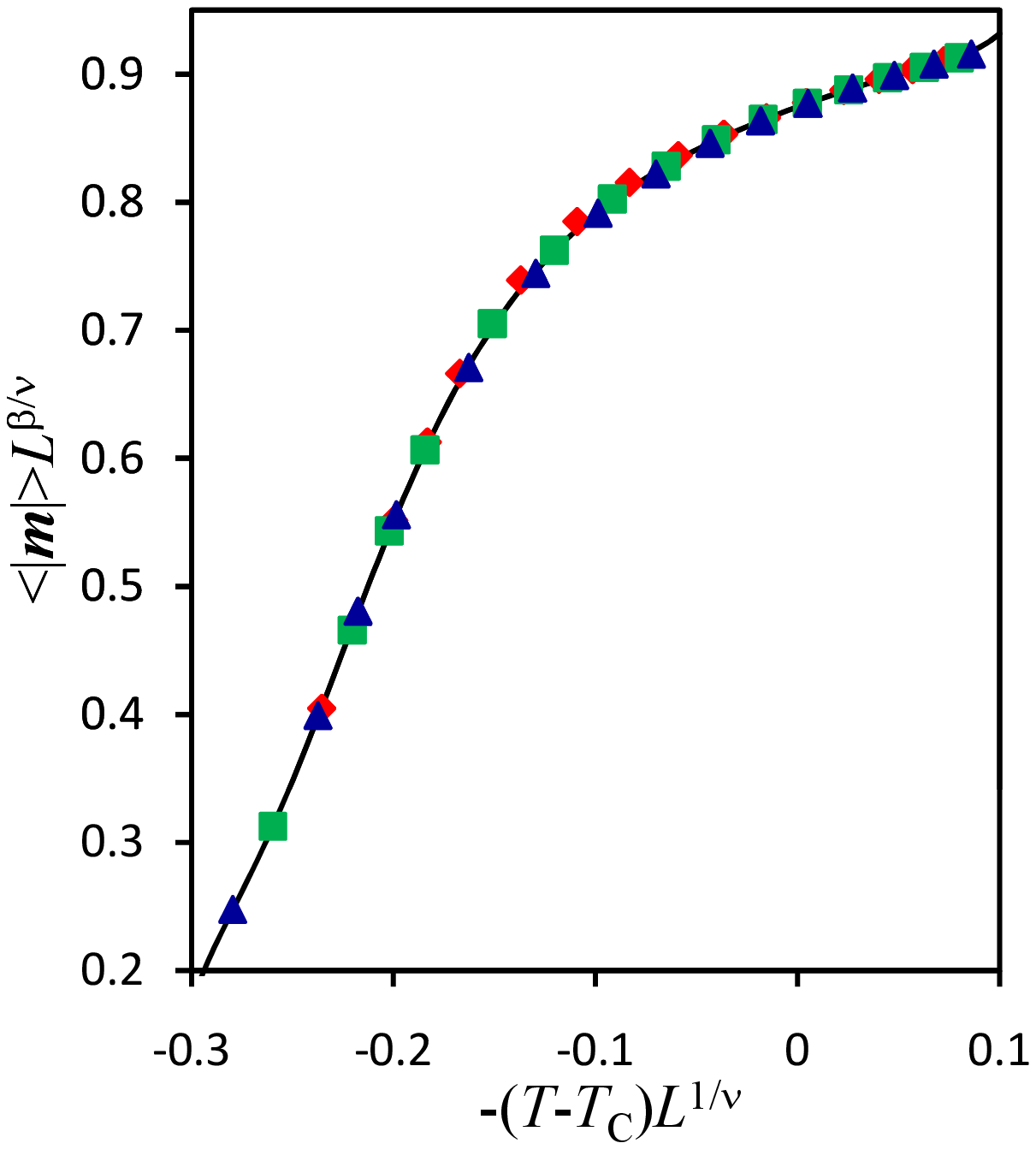}\includegraphics[width=2.5in]{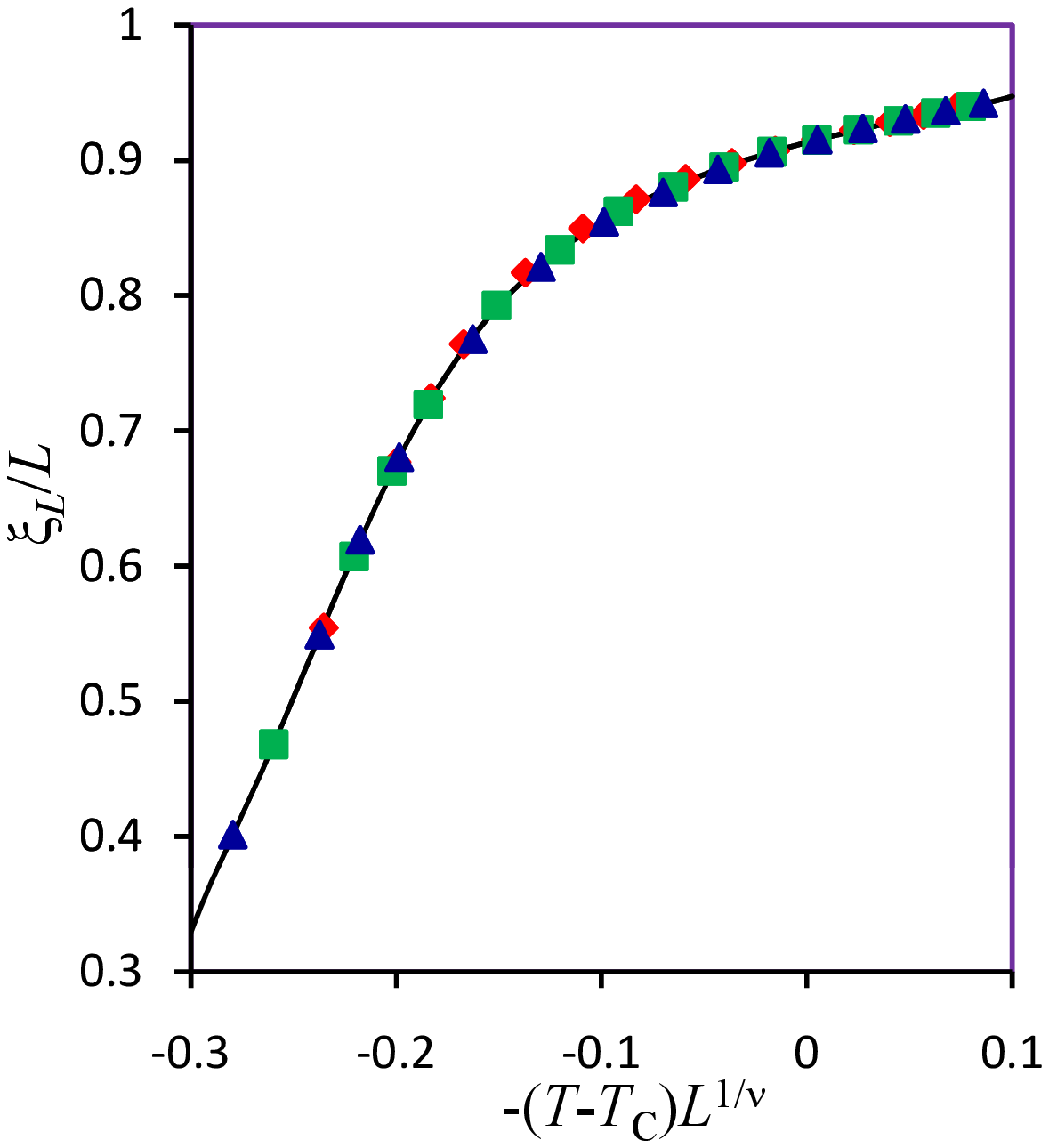}
                                 \caption{SU(2) scaling collapse plots for $U$, $\chi$, $<|\vec{m}|>$, and $\xi _L /L$.}
          \label{fig13}
       \end{figure}
Once $\gamma / \nu$ has been determined, a correlation length can be defined 
from the bare susceptibility as above (Eqn. 2). Plots of $\xi _L / L$ should cross at an infinite
lattice critical point.  Such a plot is given in Fig.~12, which shows a rather clear crossing, also around $\beta = 3.2$.
These crossings are verified by even larger confidence intervals than for $U$, with the $24^4$ data exceeding 
the $20^4$ by $18\sigma$ and the $16^4$ by $32\sigma$ at $\beta=3.5$.
Finally a full four-quantity collapse fit for $\beta _c$ and $\nu$ was performed 
utilizing $U$, $\chi $, $<|\vec{m}|>$, and $\xi _L /L$ (Fig.~13). This yields $\beta _c = 3.18(8)$ and 
$\nu = 3.42 (+0.32 -0.23)$, with a 
$\chi ^2 /d.f = 3.2$.  Again, the $16^4$ data were excluded from the fit.  The slightly high $\chi^2 / d.f.$ probably
indicates the presence of some higher order corrections to scaling still operating at these lattice sizes.  Errors 
in critical exponents and $\beta _c$ were 
determined by forcing the given quantity higher or lower until the $\chi^2 / d.f.$ doubled, with the other parameters free
to change.  This somewhat more conservative method than ``adding unity" to $\chi ^2 /d.f.$ guards against any 
possible underestimate
of statistical errors, which were from plateaus of binned fluctuations.

\section{Fits to gauge invariant quantities}

When one thinks of confinement, the quantity that immediately comes to mind is the string tension. In particular, the inverse
square root of the string tension defines a correlation length.  If the phase transition seen above is to be consistent,
the correlation length in the confining phase should match that determined from the string tension.  If the string tension
scales with the same exponent given above, that would help corroborate the existence of the phase transition, since the
string tension is determined from gauge invariant Wilson loops and does not involve any sort of gauge fixing. One can also
attempt fit the ``singular part" of the specific heat in this region to the expected scaling behavior.  Finally, the
positions of the deconfinement transition observed on lattices with one short dimension, usually interpreted as a 
finite temperature transition can be analyzed as finite-lattice $\beta$ -shifts of the infinite lattice transition seen above,
which also depend on the exponent $\nu$.  It will be shown that all three of these gauge invariant quantities scale 
consistently with the above transition seen at $\beta _{c}  = 3.18$ with $\nu = 3.4$.
  
However, first the three-dimensional scaling of the hyperlayers needs to be reconciled with the four-dimensionality of the
original system. For instance what value of $d$ should be used in the hyperscaling relationship for the specific heat
exponent $\alpha = 2-d \nu$?  The key is to consider the relationship between the correlation length being measured 
from the spin correlations in the 3-d hyperlayers
and that of the four-dimensional theory (being measured by the string tension for instance). The magnetization correlation
function can be considered similar to the 1xN Wilson loop or, more accurately, a unit 
length partial Polyakov loop(PPL) or ``gaugeon\cite{mp}."
In the Coulomb gauge a segment of links of any length, $aT$ in the fourth direction is an observable. Here $T$
is the time extent in units of the lattice spacing.  The correlation function
between two PPL's a distance $Ra$ apart is expected to behave as 
\begin{equation}
exp(-\sigma R T a^2 )
\end{equation}
where $\sigma$ becomes the usual asymptotic string tension for large $T$\cite{mp}. For $T=1$, $\sigma $
will differ 
somewhat but the dimensionality
will still be the same. The $\sigma$ for $T=1$, sometimes referred to as the Coulomb string tension,
has been observed to approximately scale with the asymptotic 
string tension \cite{greensite}.
It is therefore an inverse length-squared object. 
However, this same object was treated as the inverse correlation length itself
in the 3-d analysis, because there the 4-d links were treated as ``spins" with no inherent length associated with them, whereas in
the 4-d theory they are associated with the lattice spacing.
Therefore the correlation length determined from the spin analysis is really an area in the 4-d theory.  Thus the correct
4-dimensional exponent $\nu _4$ for a true length, $1/\sqrt{\sigma}$, should be half of the 3-d measured value of 3.42, i.e. 
$\nu _4 = 1.71$.

Neither the string tension nor the specific heat show much finite-size variation in the $\beta$ region in which they
have been measured (the region where the correlation length is less than $L$). Therefore, they can be expected to rather
closely follow infinite-lattice scaling laws. Fig.~14 shows a 
plot of $\sqrt{\sigma a^2}$ vs. $\beta$ with a one-parameter fit to the form $c(1/\beta - 1/\beta _c )^{\nu _4}$.
String tensions were taken from refs \cite{iqp,st}. A somewhat
reasonable, though not entirely satisfactory, 
fit is obtained for the values $\beta _c = 3.18$, $\nu _4 =1.71$ obtained above from the 3-d magnetization. The fit particularly
differs with the $\beta = 2.85$ point(6 standard deviations off). Without 
this point the $\chi^2 /d.f. = 3.5$. If $\nu$ is allowed to vary, only a slightly better fit is obtained for
$\nu _4= 1.67$ ($\chi ^2/d.f. = 4.08$ - fit improvement does not compensate for increasing d.f. by 1). 
Therefore the string tension appears to be selecting essentially 
the same value for the critical exponent as obtained from the Coulomb gauge magnetization.
Since string tensions are measured
with a variety of methods and on different lattice sizes, some systematic differences between the datapoints may be present,
which could explain some of the difficulties with the fit.
Similar fits can be made for a fairly broad range of $\beta _c $, so the string tension is 
not good at pinning that down further.
\begin{figure}\centering
                      \includegraphics[width=3.5in]{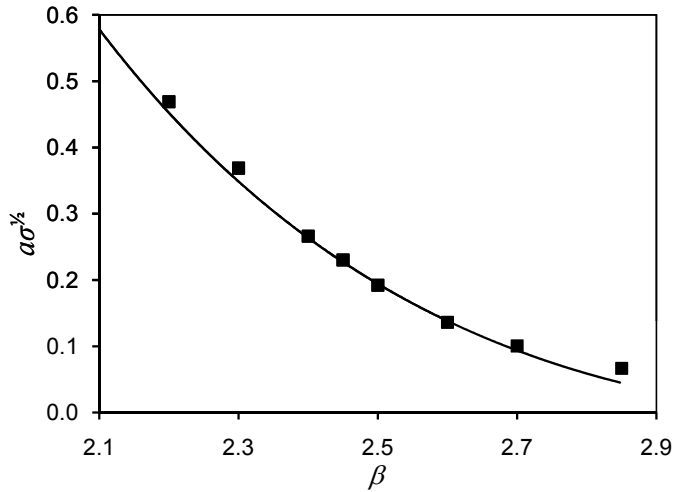}
                                 \caption{String tension fit to finite-order scaling law.}
          \label{fig14}
       \end{figure}

The specific heat exponent is expected from hyperscaling to be $\alpha = 2-4\nu _4 = -4.84$. The high negative value indicates 
a very soft singularity. 
Five derivatives of the specific heat would have to be taken before an infinite singularity would be seen.
Such a weak singularity would be virtually impossible to detect directly using numerical methods and certainly 
explains how this transition could have been missed before.  Such high values of $\nu$
are unusual in spin models, but more common in spin glasses, because the lower critical dimension(lcd) of many spin glasses 
is thought 
to lie around 2.5\cite{sghin}, and $\nu \rightarrow \infty$ at the lcd. Since it is 3-d hyperlayers that are 
being studied here, and a spin-glass to
ferromagnet transition is being hypothesized, the closeness of the dimensionality to the lcd 
could explain the unusually high value of $\nu$.  For a finite singularity, scaling can still be checked.  
Since the finite size
dependence of the specific heat is known to be very small, one can fit to the expected infinite lattice scaling function.
The following fitting function was used:
\begin{equation}
C_0 (1/\beta - 1/\beta _c)^{(4\nu _4 -2)}+C_1 +C_2 /\beta +C_3 /\beta^2 +C_4 /\beta^3 .
\end{equation}
The last four terms are needed to fit the non-singular part. $C_1$ is set to the perturbative value of 0.75, but $C_2$
through $C_4$ are allowed to differ from the perturbative values to account for higher-order 
neglected terms. Fig.~15 shows that data in the range $2.44\leq \beta \leq 3.2$ fits 
well to this form
with the previously determined $\nu _4 = 1.71$ ($\chi^2 /d.f. = 0.58$). Data were from 
plaquette fluctuations on $24^4$ lattice for runs of 500,000 sweeps or more. Again if $\nu _4$ is 
allowed to vary, the best fit
is found at a nearby value of 1.63 ($\chi^2 /d.f. = 0.57$). Without the $1/\beta ^3$ term a value of 1.53
is preferred ($\chi^2 /d.f. = 0.62)$  The specific heat is not very sensitive to the 
exact value of $\beta _c$.
\begin{figure}\centering
                      \includegraphics[width=3.5in]{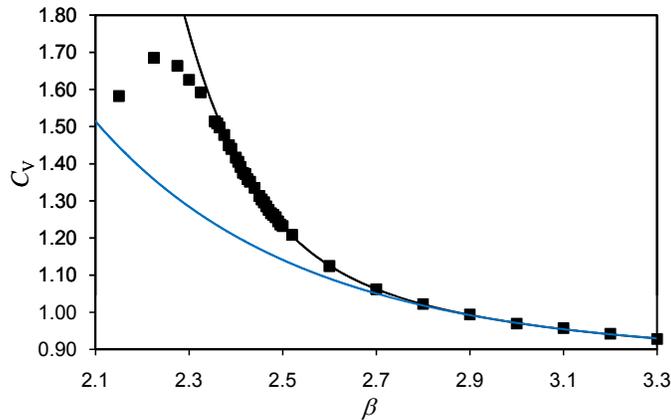}
                                 \caption{Specific heat finite-order scaling fit. Lower curve is non-singular part of fit.}
          \label{fig15}
       \end{figure}
 
Finally one can look at the scaling of $\beta_c$ on lattices with one short dimension - what has usually been interpreted
as a finite temperature deconfinement transition.  If there is an infinite-lattice critical point at $\beta = 3.2$, then
the phase transition on the distorted lattice would not be a finite-temperature transition, since the continuum limit is
already deconfined at zero temperature. Rather it is simply the same transition $\beta$-shifted by the finite size
shift effect. The shift in the critical point for a finite lattice is expected to be\cite{fss}
\begin{equation}
(\beta _c - \beta _{c\tau })\propto L_{\tau}^{-1/\nu _4}
\end{equation}
for $L_\tau << L$, where $L_{\tau}$ is the temporal lattice size.  For large $\nu$ this can be substantial and dies only
slowly with lattice size. If one plots $\beta_{c\tau} $ vs. $L_{\tau}^{-1/\nu _4}$, with $\nu _4 = 1.71$ (Fig.~16), a reasonable linear fit
is seen to Wilson-action data from Ref. \cite{pl},
with an intercept giving $\beta _c  = 3.11(4)$, close to the previously identified value of 3.18(8). Also shown is
the two-loop weak-coupling renormalization group prediction for the no phase transition hypothesis, which does not fit 
the data
as well. In addition the same quantity but for an alternative action which suppresses $Z_2$ monopoles, as studied by
Gavai\cite{gvz2}, is plotted.  These data also fit well to a linear fit with an intercept $\beta _c  = 3.13(4)$,
in agreement with
the Wilson value, and do not fit at all well with the perturbative renormalization group. 
The monopole 
suppressed action has the same weak coupling perturbation series 
as the Wilson action. For weak couplings, results should merge with
the Wilson action ones, so the $\beta _c$ values would be expected to be close, though not necessarily identical.
The system with $L\rightarrow \infty$ with $L_\tau$ finite is, of course, a true 3-d system and thus will have different critical 
exponents from the 4-d theory.  As $L_\tau \rightarrow L$, a dimensional crossover is to be expected with critical
exponents morphing to their 4-d values\cite{fisher}.
\begin{figure}\centering
                      \includegraphics[width=3.5in]{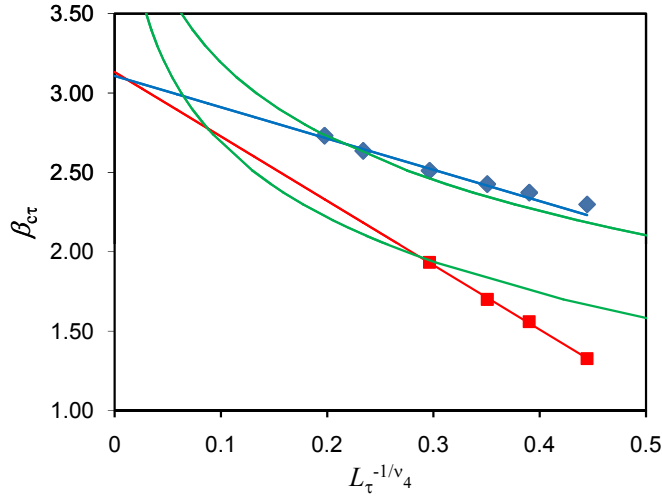}
                                 \caption{Scaling fit for finite-temperature transition points according to the 
finite-order zero-temperature
phase transition hypothesis(straight lines). Diamonds are for the Wilson action and squares are a Z2 monopole suppressed
action. Curves are the two-loop renormalization group prediction of the no phase transition hypothesis, made to pass through the 
highest $L _\tau$ points.}
          \label{fig16}
       \end{figure}

Therefore, the gauge invariant quantities of string tension, specific heat, and finite-lattice shift all scale reasonably
in accordance
with the zero temperature phase transition seen by the Coulomb gauge magnetization order parameter.  This lends further
credence to the existence of this phase transition - in particular that gauge fixing is not necessary to see it. It is
merely a very convenient and probably necessary step to take in order to define a local order parameter.

The idea that the pure gauge SU(2) lattice gauge theory might have a lattice-artifact driven phase transition in
analogy to U(1) has been considered before\cite{zp,ps}.  Perhaps the best previous evidence to date that 
such a transition must exist
involves another expansion of the coupling space, the fundamental-adjoint plane.  For a sufficiently large adjoint coupling
a first order bulk transition appears in the SU(2) theory.  The scaling of latent heat with the size
of the hysteresis region was shown in Ref. \cite{fa} to be inconsistent with the endpoint of the first order line being
an ordinary critical point.  Rather it is only consistent with being a tri-critical point associated with an exactly broken
symmetry, for which the transition
cannot just end, but must continue as a higher-order one, again until it hits an edge of the phase diagram. Slopes of
lines of constant physics predict a crossing of the Wilson axis. Consistent with this hypothesis, when ``finite temperature"
transition points are followed into the fundamental-adjoint plane they appear to head toward this 
tricritical point\cite{ggm}. Exact joining is difficult to establish, however, and has been disputed\cite{blum}, but
if the endpoint of the first-order line is truly a tricritical point then joining must take place.

\section{Conclusion}
It has often proven useful to connect theories by using an extended coupling space, especially when the behavior of one
theory is better known than the other.  This paper extends an old suggestion 
by Darhuus and Fr\"{o}hlich relating the 4-d SU(2) lattice gauge theory
to the 3-d O(4) Heisenberg model\cite{frohlich}, by allowing the couplings of horizontal and vertical plaquettes to differ.
The enlarged model becomes the 3-d O(4) Heisenberg model in the $\beta _H \rightarrow \infty$ limit and the 4-d SU(2) lattice
gauge theory in the $\beta _H =\beta _V$ limit. A Monte Carlo simulation at $\beta _H = 20$ shows a phase transition
at $\beta _{Vc} = 1.01$. In order to follow the same ferromagnetic
order parameter as in the spin model, the lattice gauge theory must be transformed into Coulomb gauge, 
in which the 4th direction pointing links
act as O(4) spins exhibiting a 3-d global symmetry, the remnant symmetry of the gauge fixing.  Although Coulomb gauge fixing has 
been plagued with large systematic and random errors introduced by the gauge fixing algorithm, it is found that
using open boundary conditions solves these problems - with errors hundreds of times less than when using periodic boundary
conditions, nearly eliminating the lattice Gribov problem. The local order parameter allows 
standard methods for finding phase transitions such as Binder cumulant crossings and finite-size scaling collapse fits.
Not surprisingly, because of the high value of $\beta _H$, the $\beta_H = 20$ transition is very 
similar to the Heisenberg transition at $\beta _H = \infty$, $\beta _V = 0.936$, with similar critical exponents.  However,
the fact that the transition still exists for non-infinite $\beta _H$ has far-reaching implications.  
The ferromagnetic transition breaks an exact symmetry,
the remnant gauge symmetry left over after Coulomb gauge fixing.  Therefore, it must divide the coupling plane into two
distinct regions - symmetry broken and unbroken.  The line of phase transitions must persist until it hits another edge.  
The Heisenberg model itself has only one critical point, so the line of phase transitions must cross the line $\beta _V =
\beta _H$ at a finite value (somewhere between 1.01 and 20 - as $\beta_H$ decreases, $\beta _{Vc}$ increases).
Indeed a high-order transition is seen around $\beta = 3.2$, showing Binder cumulant and $\xi _L /L$ crossings, and good
collapse of various graphs to scaling functions about this point.  The transition also predicts scaling laws 
consistent with the  
behavior of the specific heat, string-tension, and even finite-temperature transition points (reinterpreted as 
finite-lattice shifts) in the crossover region.  

Much of the lattice gauge theory program has been based on the assumption that the non-abelian gauge theories SU(2)
and SU(3) have no phase transition and the continuum limit is confining.  Of course the SU(3) case needs to be 
checked in detail, but if SU(2) has a zero-temperature phase transition it is likely SU(3) has one as well.  A non-confining
and symmetry-broken continuum limit would require a new mechanism of quark confinement, since it would no longer be 
a property of the gluon sector acting alone.  One possibility is that the light-quark chiral condensate is itself the
source of confinement, a hypothesis which has been advanced by Gribov\cite{gribov} and others\cite{csb}.  
A strong enough force, whether confining or not can cause quark-antiquark pairs to condense in 
the vacuum, breaking chiral symmetry.  If this chiral vacuum expels strong color fields, of which there is some 
evidence\cite{markum}, then the chiral vacuum itself could cause a bag-like pressure around strong color sources, with the
lowest energy configuration of vacuum + hadron being one in which the color fields are confined to a small volume. In
most cases unbreaking of chiral symmetry from thermal effects and deconfinement are coincident, 
which supports such a linkage.
Thus it is possible that simulations with light quarks are exploring correct continuum physics (whereas the pure-glue
simulations in the confining phase 
are exploring a strong-coupling phase not connected to the continuum limit). To be sure, alternative gluon 
actions
need to be explored which eliminate, if possible whatever lattice artifacts are causing the pure-glue 
phase transition, in analogy
with the monopoles of the U(1) theory.  Preliminary data suggest that eliminating both negative plaquettes (Z2 vortices
and monopoles)
{\em and} a different kind of monopole defined from the non-abelian Bianchi identity, which I have called SO(3)-Z2
monopoles\cite{so3-z2}, may force the SU(2) theory to remain in the ferromagnetic phase for all couplings.

The symmetry breaking which occurs in the weak coupling phase is not unexpected.  The U(1) theory in the 
Coulomb gauge almost certainly is also magnetized in the same way, becoming paramagnetic as the coupling becomes
stronger at the monopole-induced
phase transition to a confining theory, but being magnetized in the weak-coupling region 
down to the weak-coupling continuum limit.
In this gauge, at least, the photon is a spin-wave like Goldstone boson associated with
the breaking of the remnant symmetry left over after Coulomb gauge fixing, as opposed to the original bare gauge field
the theory started out with.  This picture has been previously explored in the continuum\cite{photongoldstone}. 
The Gupta-Bleuer quantization method may be an alternative way to see how the elementary fields build into a Goldstone
boson, and how a linked gauge and Lorentz symmetry is ``reborn."  A similar picture may hold for the pure-gauge
SU(2) and SU(3) gauge theories, with the gauge fields remaining massless and the 
running coupling having an infrared fixed point. However, when light quarks are added, this vacuum could become unstable
and confinement may emerge as a byproduct of chiral symmetry breaking, following Gribov's or a similar mechanism.

\section*{Acknowledgement}
David J. Starling contributed to an earlier version of this work.

\end{document}